\documentclass[prl,twocolumn,showpacs,preprintnumbers,amsmath,amssymb,revtex4-1]{revtex4-1}
\usepackage{amsfonts, amsmath, bm, bbm, graphicx, epsfig, epstopdf}
\usepackage{dcolumn}
\usepackage{textcomp}
 \usepackage[caption=false]{subfig}
 \usepackage[caption=false]{subfig}

\usepackage{xcolor}
    \usepackage{braket}
\renewcommand\bra[1]{{\langle{#1}|}}
\makeatletter
\renewcommand\ket[1]{%
  \@ifnextchar\bra{\k@t{#1}\!}{\k@t{#1}}%
}
\newcommand\k@t[1]{{|{#1}\rangle}}
\makeatother

\begin{document}
\title{Electron cloud design for Rydberg multi-qubit gates}

\author{Mohammadsadegh Khazali}
\affiliation{Institute for Quantum Optics and Quantum Information of the Austrian Academy of Sciences, A-6020 Innsbruck, Austria}
\affiliation{School of Physics, Institute for Research in Fundamental Sciences (IPM), Tehran 19395-5531, Iran}

\author{Wolfgang Lechner}
\affiliation{Institute for Theoretical Physics, University of Innsbruck, A-6020 Innsbruck, Austria}
\affiliation{Parity Quantum Computing GmbH, A-6020 Innsbruck, Austria}

\date{\today}
\begin{abstract}
This article proposes quantum processing in an optical lattice, using Rydberg electron's Fermi scattering from ground-state atoms in spin-dependent lattices as a source of interaction. Instead of relying on Rydberg pair potentials, the interaction is controlled by engineering the electron cloud of a sole Rydberg atom. Here we specifically propose the implementation of two prominent multi-qubit gates i.e. the stabilizer-phase operator and the Toffoli gate. The new scheme addresses the main bottleneck in Rydberg quantum simulation by suppressing the population of short-lived Rydberg states over multi-qubit operations. This scheme mitigates different competing infidelity criteria, eliminates unwanted cross-talks, and allows operations in dense atomic lattices. The restoring forces in the molecule type Ryd-Fermi potential preserve the trapping over a long interaction period.  The features in the new scheme are of special interest for the implementation of quantum optimization and error correction algorithms.
 \end{abstract}

\maketitle

The recent experimental progress in the development of highly controllable quantum systems using Rydberg atoms enabled a parameter regime where fully functional quantum information processing algorithms are in reach. While individual control of atoms and their interactions has been achieved, the next step is the development of complex interactions between multiple quantum bits (qubits) which are of great interest for applications ranging from quantum optimization to error-correcting codes. 

A particular application of near-term quantum devices is solving optimization problems using a gate model approach. The parity architecture \cite{LHZ15,End21},  translates a problem with all to all connectivity to a simple nearest-neighbor problem-independent interaction.
Hence, the quantum approximate optimization algorithm (QAOA) \cite{Far14,Far16} implementation would be simplified to programming single-qubit operations, as well as applying problem independent four-body stabilizer-phase gate. 
The other prominent multi-qubit gate is Toffoli, playing a pivotal role in quantum error correction \cite{Cor98,Sch11}, fault tolerant quantum computation (QC) \cite{Den01,Pae13} and Shore's algorithm \cite{Sho95}. 

The realization of multi-qubit gates with the concatenation of one- and two-qubit gates \cite{Mot04,Lec20} results in a significant overhead \cite{Mas03,She09}.
The long-range many-body Rydberg interaction is vastly used in different quantum operations \cite{Saf10,Ada19,Jak00,Bro20,Ise10,Wil10,Gra19,Kha19,Gr017,khaz2020rydberg,Kha16,KazRev} including direct operation of multi-qubit gates \cite{Shi18,Bet18,Lev19,Kha20,Ise11,Gla17}.
The fidelities of these direct Rydberg multi-qubit gates are limited by the competing requirement of the presence (absence) of inter- (intra-) component interaction in addition to the short lifetime of the Rydberg level.

This article  introduces a versatile toolbox to engineer multi-qubit operations by designing the Rydberg electronic cloud with respect to the lattice geometry.
A spin-dependent lattice \cite{Dut98,Sol11,Kar09,Lee07,Jak99,Bre99,Bri00,Man03,Mandel03,Kum18} is used to harvest the Fermi-scattering of the Rydberg electron \cite{Gre00,Gaj14} from neighboring lattice sites as a source of coherent interaction for the desired gates.
The spin-dependent geometrical shift of  atoms accommodates them inside or outside of  the Rydberg electron's wave-function,  providing a spin-dependent nearest-neighbor interaction appropriate for gate operations.  
This double-encoding of qubits in internal and external degrees of freedom distinguishes the gate from previous proposals without spatial encoding of the qubit. 
In a previous work, spin-flip via Rydberg-Fermi interaction at very short inter-atomic distances below 50nm has been studied  \cite{Nie16}. 
However such short lattice constants are not realizable in the experiment. The dual spin/spatial encoding proposed in this article  provides strong qubit-dependent interaction within the present optical lattices.

In terms of scalability, making dense atomic lattices would be a promising approach in passing from the current era of intermediate-scale quantum devices.
The new advances in sub-wavelength imaging and laser addressing of atoms \cite{Aga06,Cho07,Kap10,Wei11,Mil13,Yav09,Vei21,Sub19} allow operation precisions in the nanometer scale. %with the new record of high resolution imaging bellow 200nm \cite{Vei21}, 
These advances call for new techniques for multi-qubit interactions at ultra-short distances. In the Rydberg-Fermi scheme, the interaction to loss ratio improves by going to smaller lattice constants, with examples studied here at 400nm and 500nm lattice constants. The restoring forces in the molecule type Rydberg-Fermi potential preserve the trapping over a long interaction period. 
The other bottleneck in the scalability of the Rydberg quantum processors comes from the lifetime of the Rydberg states. At small inter-atomic distances, the dipolar scheme is limited to short-lived Rydberg states with low principal number to avoid strong level-mixing and line-broadening \cite{Kil04,Kea13,Bij15}. In the Rydberg-Fermi scheme, the absence of strong level-mixing close to the outer shell of the Rydberg wave-function, allows for choosing highly excited Rydberg states with longer lifetime.
  Furthermore, the Rydberg-Fermi gate schemes, significantly reduce the Rydberg population over the gate operation time. For example in C$_k$-NOT, the Rydberg population averaged over $2^{k+1}$ qubit configurations scales by $k$ in dipolar schemes \cite{Ise11,Kha20}. In contrast in the Rydberg-Fermi protocol the Rydberg population scale $2^{-(k+1)}$ is in favour of multi-qubit operations, see App.~C. This would also be of special interest for quantum search algorithm \cite{Mol11,Pet16}.
  
The Rydberg-Fermi scheme is especially appealing for {\it fast} multi-qubit operations. In an architecture of one vrs many qubits, the presence/absence of inter/intra component interaction paves the way for the implementation of C$_k$-NOT \cite{Ise11,Bet18,Kha20,Shi18} and C-NOT$^k$ \cite{Kha20} gates in minimal pulse steps as well as direct operation in logical basis \cite{Khaz22Log}.
 In Rydberg-dipolar schemes, the fast operation is carried out by exciting all $k$ control atoms in $\ket{0_c}$ state to the Rydberg level followed by target Rydberg rotation  \cite{Ise11,Shi18,Kha20}. This results in competing requirements i.e. to preserve the lattice trapping against strong dipolar interaction, to overcome/preserve the blockade between inter/intra components, and to not excite the neighboring Rydberg levels \cite{Saf05,Ise11,Zha12,The16,Bet18,Kha20}. 
In the Rydberg-Fermi approach, multi-qubit operations are carried out by exciting a single atom to the Rydberg state. The system thus operates the many-body interaction at a different regime of energy hierarchy without the mentioned rivalry in Rydberg-dipolar systems.  
Furthermore, the absence of intra-component interaction eliminates the unwanted phase errors in multi-qubit gates, see the App.~C.

\section{Results}

\subsection{Rydberg-Fermi interaction in a qubit-dependent lattice}
\label{Sec_RydFermi}

\begin{figure}
\centering 
 \scalebox{0.48}{\includegraphics{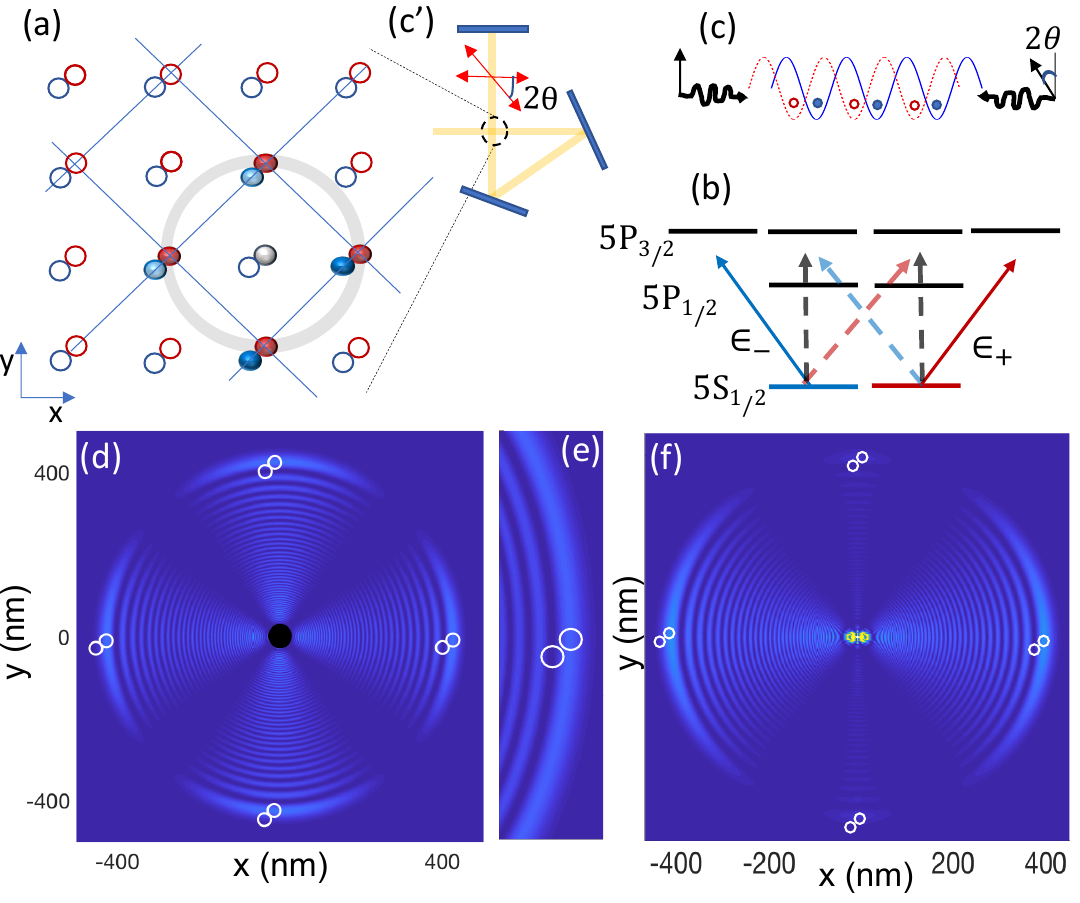}} 
\caption{Rydberg-Fermi interaction in Spin-dependent lattice. 
 (a)   In a 2D structure with a single atom per site, applying a qubit-dependent lattice-shift makes each atom in a spatial superposition of being in red and blue sites where the components are controlled by the internal electronic qubit-states $|0\rangle$ and $|1\rangle$. 
 Hence the Rydberg electron of the central atom would exclusively scatter from the plaquette atoms in a specific spin-lattice, providing qubit-dependent interaction.
(b) In $^{87}$Rb, tuning the trapping laser between  $5P_{3/2}$ and $5P_{1/2}$, the polarizability of qubit states $|0\rangle$ and $|1\rangle$ are given by distinguished circularly polarized lights $\varepsilon_{-}$ and $\varepsilon_{+}$ respectively.  (c,c') Counter propagating linearly polarized lights with relative polarization of $2\theta$,  forms two distinguished optical-lattices of $\varepsilon_{-}$ and $\varepsilon_{+}$ displaced by $D_{\{x,y\}}=2\theta/k$ in each dimension, trapping different qubit states.   
(d) Exciting the Rydberg superposition state of Eq.~\ref{Eq_Sup} with n=64 with the quantization axis being perpendicular to the lattice plane provides a symmetric interaction over the four neighboring plaquette atoms.   (e) The zoomed vision  shows  that the  $|0\rangle$ and $|1\rangle$ qubit states of the plaquette atoms are localized on the node and anti-node of the Rydberg wave-function, providing a qubit-dependent contrast of Fermi scattering.   
 (f)  Two-color excitation of $(|65P_{3/2},1/2\rangle+|65P_{1/2},1/2\rangle)/\sqrt{2}$ with in-plane quantization axis provides  couplings with two opposite plaquette atoms in the lattice. }\label{Fig_SpiLat}
\end{figure}

\begin{figure}
\centering 
       \scalebox{0.36}{\includegraphics{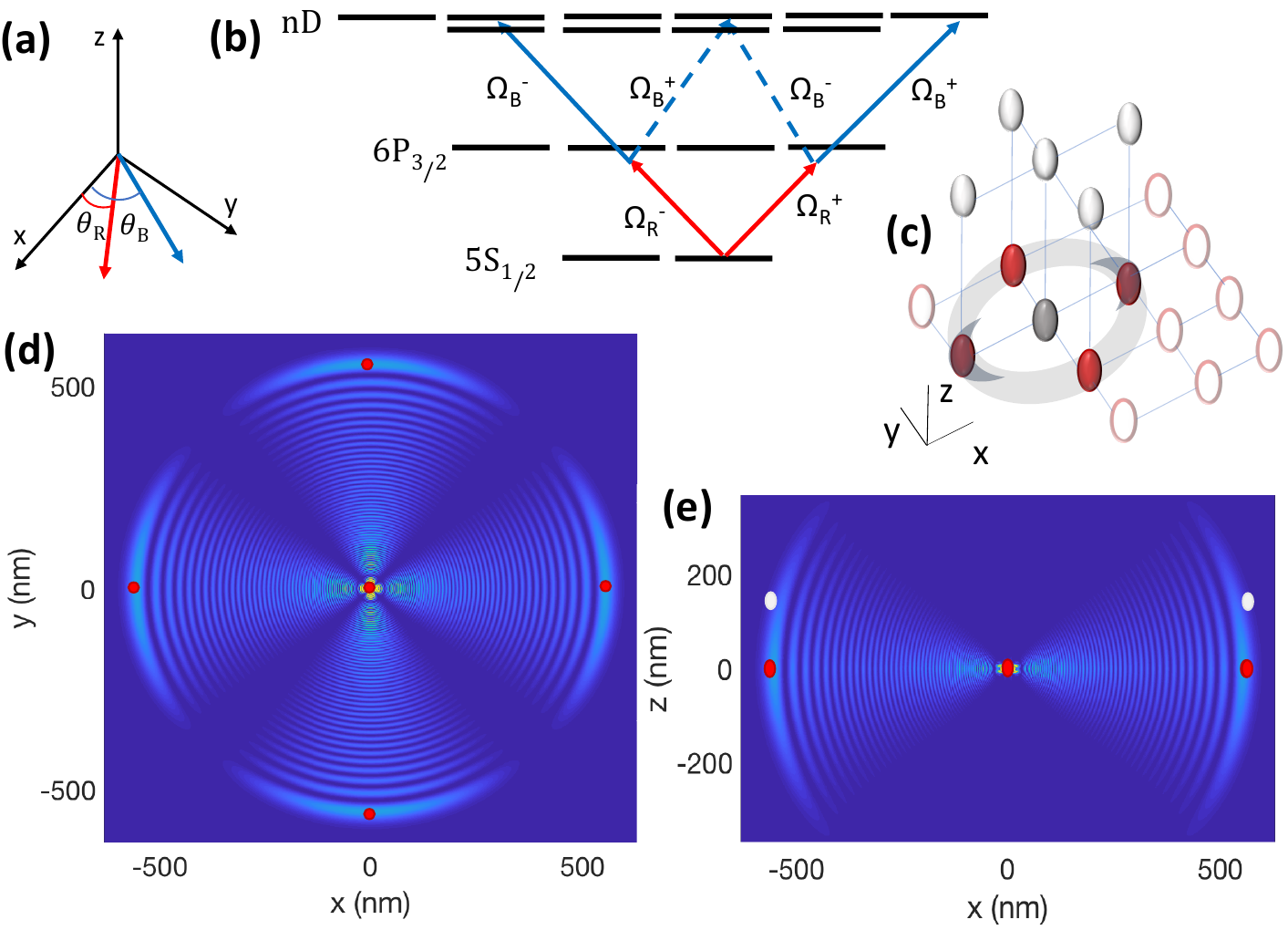}} 
        \caption{Superposition of Rydberg states. (a) The desired Rydberg superposition could be controlled by the polarization angles $\theta_{R,B}$ of the two linearly polarized fields $\Omega_{R,B}$ propagating along the $z$ direction. (b) The transitions shown by dashed-lines would form destructive interference when  $\theta_R-\theta_B=\pi/2$, leading to the superposition of $(\text{e}^{\text{i}(\theta_R+\theta_B)}|m_j=5/2\rangle+\text{e}^{-\text{i}(\theta_R+\theta_B)}|m_j=-3/2\rangle)/\sqrt{2}$ states of 64D Rydberg level. (c) Applying the qubit-dependent shift perpendicular to the lattice plane allows tunning the in-plain inter-atomic distance.  The (d) $xy$ and (e) $xz$ cross-sections of Rydberg-Fermi interaction. The excited Rydberg superposition state would be further confined around the position of plaquette atoms and hence enhances the interaction. Red and White ovals present the qubit-dependent position of $|0\rangle$ and $|1\rangle$ states.
}\label{Fig_RydSuper}
\end{figure}

In a two-dimensional lattice shown in Fig.~\ref{Fig_SpiLat}a, atoms in spin-states  $\ket{0}$, $\ket{1}$ are trapped in shifted lattices distinguished by red and blue. The gate operations are carried out by exciting the central atom in a plaquette to the Rydberg level. Depending on whether the central atom is excited from  $\ket{0}$ or $\ket{1}$  state, the plaquette atoms in  $\ket{0}$ or $\ket{1}$ spin-lattice would be localized on the nodes and antinodes of the Rydberg electron's last lobe, see Fig.~\ref{Fig_SpiLat}d-\ref{Fig_SpiLat}f. This provides contrast on the Fermi scattering of the electron from distinguished qubit states of plaquette atoms. 
The qubit-dependent interaction could also be realized by the spin-dependent shift perpendicular to the lattice plane as depicted in Fig.~\ref{Fig_RydSuper}c.

The {\it spin-dependent lattice} is formed by counter-propagating linearly polarized lights, see Fig.~\ref{Fig_SpiLat}b,c.
Introducing a relative shift between the fields' polarizations of $2\theta$, the total electric field can be written in terms of the sum of right and left circularly polarized lights $E=E_0 \exp(-i \nu t)(\varepsilon_{+}\sin(kz+\theta)+\varepsilon_{-}\sin(kz-\theta))$. 
To make a spin-dependent lattice-shift, the spin polarizabilities should be linked to different circular polarization components of lights \cite{Dut98}. To cancel the polarizabilities with unwanted light elements shown by dashed lines in Fig.~\ref{Fig_SpiLat}b, the trapping laser must be tuned between $P_{3/2}$ and $P_{1/2}$ states so that the ac-Stark shifts of these two levels cancel each other.
As a result the $m_j=\pm1/2$ levels of the ground state would be trapped by $V_{\pm}=\alpha |E_0|^2 \sin(kz\pm\theta)$.
The hyperfine qubit states $|0\rangle=|F=1,m_F=1\rangle$ and  $|1\rangle=|F=2,m_F=2\rangle$ experience $V_0=(V_{+}+3V_{-})/4$ and $V_{\ket{1}}=V_{+}$.
A spin-dependent lattice provides dual spin/spatial encoding of the qubit. 
A Raman transition coherently transfers atoms from one internal state to the other, thereby causing hopping between the two Wannier-functions \cite{Jak98,Jak03,Maz10}. The spin rotation Rabi frequency in qubit-dependent lattice would be modified by the  Frank-Condon factor, see  App.~B.

{\it Ryd-Fermi Interaction --} The interaction between the Rydberg electron and the ground state atom is a Fermi type pseudo potential \cite{Fer,Eil17,Eil19},
\begin{equation}
\label{Eq_RydFermi}
V_{\text{RF}}=(2\pi \frac{\tan(\delta^s)}{k(R)}-6\pi\frac{\tan(\delta^p)}{k^3(R)}\stackrel{\leftarrow}{\nabla}_{{\bf r}}.\stackrel{\rightarrow}{\nabla}_{{\bf r}})\delta({\bf r}-{\bf R})
\end{equation}
with {\bf r} and {\bf R} being the positions of the Rydberg electron and the ground state atom with respect to the ionic core, and $\delta^{\{s,p\}}$  are the triplet s- and p-wave scattering phase shift of the Rydberg electron from the neighboring ground state atom \cite{khu02}. 
The electron wave-vector $k(R)$ is defined by the kinetic energy of the Rydberg electron at energy $E=-1/2n^2$ when it collides with a ground-state atom, i.e. $k^2(R)/2=E+1/R$.
The level-shift caused by the Rydberg electron scattering from the $l^{\text{th}}$ plaquette atom in the qubit state $i=\ket{0,1}$ would be characterized by two parameters 
 \begin{eqnarray}
 \label{Eq_V01}
&&\bar{V}_{\text{RF}\ket{i_l}}=\int |w({\bf R}-{\bf l_{i}})|^2 V_{RF}({\bf R}) \text{d}{\bf R}\\ \nonumber
&&\text{MD}_{V_{\text{RF}\ket{i_l}}}={\int |w({\bf R}-{\bf l_{i}})|^2 |V_{RF}({\bf R})-\bar{V}_{\text{RF}\ket{1_l}}| \text{d}{\bf R}}
\end{eqnarray} 
that are the  average and the mean deviation of the scattering energy over the $l^{th}$ plaquette atom's  Wannier-state $w$ centred at $l_i$.
Considering the symmetry of the Rydberg wave-function all plaquette atoms would experience the same qubit-dependent interactions. 
The Rydberg electron's wave-packet dynamics are in the ps range \cite{Ger91,Gar95}. Therefore, over the MHz scale of operation, the interaction of the Rydberg electron with all the plaquette atoms in $\ket{1}$ qubit state would be alike and add up, see \cite{Gaj14}.

 Two-color excitation of the superposition state $(|65P_{3/2},1/2\rangle+|65P_{1/2},1/2\rangle)/\sqrt{2}$ with in-plane quantization axis provides sites' specific couplings along a line, see Fig.~\ref{Fig_SpiLat}f. The two-color light could be obtained in a setup of beamsplitters and acusto-optical modulators. The generated superposition mainly contains the $Y_{1,0}$ spherical harmonic term, which significantly concentrates the electron wave-function along the quantization axis and hence enhances the interaction strength.  The atoms prepared in ground motional state \cite{Kau12,Mad20,Tho13,Bel13} are considered delocalized over the Gaussian wave-function. 
 Hence they would experience an effective Rydberg-Fermi interaction that is averaged over their spatial profile. 
The scattering energy of Rydberg electron over the qubit-dependent Wannier state of  the l$^{th}$ plaquette atom with FWHM=20nm would be quantified by Eq.~\ref{Eq_V01} as
\begin{eqnarray}
\label{Eq_64D}
 &&\bar{V}_{\text{RF}\ket{1_l}}= 2\text{MHz},\quad \text{MD}_{V_{\text{RF}\ket{1_l}}}=0.1\text{MHz}\\ \nonumber
 &&\bar{V}_{\text{RF}\ket{0_l}}=0.3\text{MHz}, \quad \text{MD}_{V_{\text{RF}\ket{0_l}}}=0.2\text{MHz} \quad \quad
 \end{eqnarray}
where the qubit-dependent lattice-shift of $D=36.8$nm is considered.

To apply a uniform interaction on all the plaquette atoms, the polarization axis must be perpendicular to the lattice plane, see Fig.~\ref{Fig_SpiLat}d and \ref{Fig_RydSuper}d.
The interaction enhancement can be obtained by exciting {\it superposition of Rydberg levels}. The spatial constructive (destructive) interference of Rydberg wave-functions over the position of plaquette qubits (elsewhere) could further confine the electron and hence enhances the interaction. 
The desired Rydberg superposition could be controlled by the polarization angles $\theta_{\{R,B\}}$ of the two linearly polarized lights $\Omega_{\{R,B\}}$ used for Rydberg excitations, see Fig.~\ref{Fig_RydSuper}. These fields are propagating perpendicular to the lattice plane along the $z$ direction. 
The linear polarized light could be expressed in terms of circular polarizations  $\Omega_{j}=(\exp(\text{i}\theta_j)\Omega_j^{+}+\exp(-\text{i}\theta_j)\Omega_j^{-})/\sqrt{2}$. Adjusting $\theta_R-\theta_B=\pi/2$ the transition to $\ket{nD_{j},m_j=1/2}$ would be canceled by destructive interference.  Hence the excited state would be 
\begin{equation}
\label{Eq_Sup}
e^{\text{i}(\theta_R+\theta_B)} \frac{\ket{nD_{\frac{5}{2}},\frac{5}{2}}}{\sqrt{2}}+\text{e}^{-\text{i}(\theta_R+\theta_B)} \frac{\ket{nD_{\frac{5}{2}},\frac{-3}{2}}+\ket{nD_{\frac{3}{2}},\frac{-3}{2}}}{2}. 
\end{equation}
The simultaneous excitation of both $\ket{nD_{3/2}}$ and $\ket{nD_{5/2}}$ could be obtained by beam splitting and frequency adjustment of the blue laser. 
This could be realized in a setup of beamsplitters and acusto-optical modulators. 
 The polarization angles in Eq.~\ref{Eq_Sup} would act as a controlling parameter to rotate the interaction maxima e.g.  $\theta_R+\theta_B=\pi,0$ are corresponding to Fig.~\ref{Fig_RydSuper}d and the same pattern rotated by $\pi/4$ around the $\hat{z}$ axis.  For the sake of presentation, Fig. 1 and 2 only plot the s-wave scattering part of $V_{\text{RF}}$, which is the dominant term at the desired last lobe in $^{87}$Rb atoms.  The exciting laser's polarization and propagation direction could act as a controlling knob to program different interaction connectivities among neighboring lattice sites.

 In the setup of Fig.~\ref{Fig_SpiLat}d with lattice constant of 400nm, exciting the central atom to  the superposition state of Eq.~\ref{Eq_Sup} with n=64, the scattering energy of the Rydberg electron over the qubit-dependent Wannier state of  the l$^{th}$ plaquette atom with FWHM$_{x,y}$=20nm and FWHM$_{z}$=35nm  would be quantified by Eq.~\ref{Eq_V01} as
\begin{eqnarray}
\label{Eq_64Sup}
 &&\bar{V}_{\text{RF}\ket{1_l}}=2.5\text{MHz},\quad \text{MD}_{V_{\text{RF}\ket{1_l}}}=0.25\text{MHz}\\ \nonumber
 &&\bar{V}_{\text{RF}\ket{0_l}}=0.35\text{MHz}, \quad \text{MD}_{V_{\text{RF}\ket{0_l}}}=0.2\text{MHz}, \quad \quad
 \end{eqnarray}
where the  magic lattice-shift of $D=35.6$nm in the qubit-dependent structure results in uniform $\bar{V}_{\text{RF}\ket{0_l}}$ inside and outside the last lob.  
To further narrow the interaction induced line broadening $\text{MD}_{V_{\text{RF}\ket{i_l}}}$, ultra-tight confinement of atoms could be obtained by the dark-state optical-lattices \cite{Yav09} and deep confinement to sub-nanometer scale is expected in Rydberg-empowered optical-lattices \cite{Kha23}.

\begin{figure}
\centering 
       \scalebox{0.34}{\includegraphics{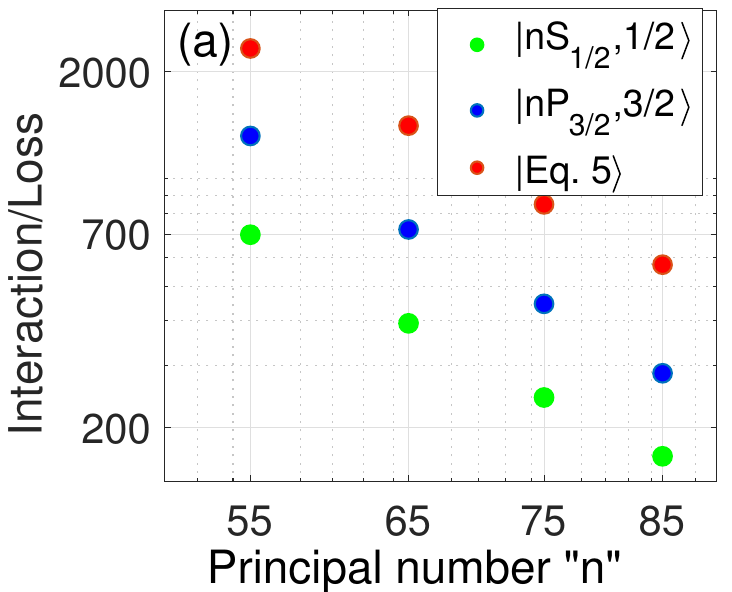}} 
       \scalebox{0.34}{\includegraphics{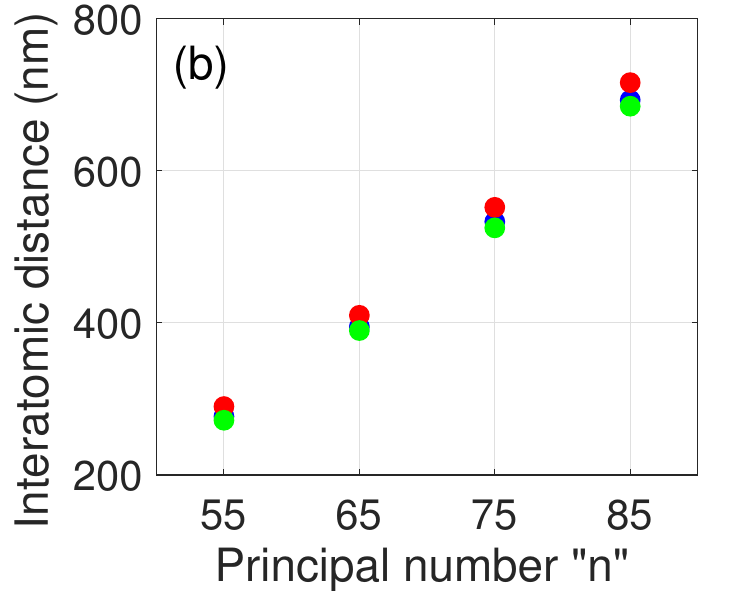}} 
\caption{The scaling of interaction to loss ratio. (a) The scattering interaction of Rydberg electron from the four plaquette atoms over the decay rate of the central Rydberg atom is plotted as a function of the principal number for  Rydberg states $|nS_{1/2},1/2\rangle$, $|nP_{3/2},3/2\rangle$, and for the state presented in Eq.~\ref{Eq_Sup}. %(b) Going to a higher principal number is associated with a larger inter-atomic distance, facilitating the single-site addressing. 
While smaller $n$ enhances the coherence, (b) the smaller inter-atomic distance raises concern about single site addressing.
}\label{Fig_Scale}
\end{figure}

The scale of interaction to loss ratio improves by going to smaller Rydberg principal numbers. 
The volume of the Rydberg electron scales by $n^6$ \cite{Eil19}. Hence the electron density and the interaction scales by $V_{\text{RF}}\propto n^{-6}$. Since the lifetime of the Rydberg state scales by $n^3$ \cite{Bet09}, the interaction to loss ratio would scale by $n^{-3}$, see Fig.~\ref{Fig_Scale}a. While the lattice configuration of Fig.~\ref{Fig_RydSuper}c allows  tuning the desired inter-atomic distance, going to a low principal numbers would raise concerns about single site addressing, see App.~E. 
The new advances in  sub-wavelength trapping, spin rotating and imaging  \cite{Yav09,Aga06,Cho07,Sub19,Kap10,Wei11,Mil13} provide a wide range of opportunities for the Rydberg-Fermi QC. Also, dual-species lattices \cite{She21,Sin21,LHZ15,Huf22} could be used to suppress the laser-cross talk issues in compact lattices.  

Applying the lattice shift perpendicular to the lattice plane provides a freedom in choosing the inter-atomic distance, see Fig.~2c-e. 
 In the $\lambda=1064$nm optical lattice \cite{Wei11}, exciting the central atom to  the superposition state of Eq.~\ref{Eq_Sup} with n=74, the scattering energy of the Rydberg electron over the qubit-dependent Wannier state of  the l$^{th}$ plaquette atom with FWHM$_{x,y}$=25nm and FWHM$_{z}$=35nm  would be quantified by Eq.~\ref{Eq_V01} as
\begin{eqnarray}
\label{Eq_75Sup}
 &&\bar{V}_{\text{RF}\ket{1_l}}=1.1\text{MHz},\quad \text{MD}_{V_{\text{RF}\ket{1_l}}}=0.07\text{MHz}\\ \nonumber
 &&\bar{V}_{\text{RF}\ket{0_l}}=0.1\text{MHz}, \quad \text{MD}_{V_{\text{RF}\ket{0_l}}}=0.05\text{MHz}, \quad \quad
 \end{eqnarray}
where the qubit-dependent lattice-shift of $D_{z}$ = 180nm is considered.

\section{Implementation of multi-qubit gates}
\label{Sec_Implementation}

\subsection{Parallelized  gate}
The  Ryd-Fermi interaction in qubit-dependent  atomic lattice could be used for the implementation of the parallelized  gate 
\begin{equation}
\label{Eq_ParalGate}
U_g=|0\rangle_c\langle 0| \otimes \mathbb{I}+|1\rangle_c\langle 1| \otimes \prod \limits_{i=1}^4 \sigma^i_x
\end{equation}
 which is an essential element in realizing the stabilizer-phase gates \cite{Wei10}, see Methods for detailed discussion.

In this proposal, the target atoms are localized on the square plaquette around the central control atom. Exciting the $\ket{1_c}$ state of the control atom to the Rydberg level, its electron creates potential energy shits via Fermi scattering. In the spin-dependent lattice, the contrast of scattering energy depends on the presence or absence of the Rydberg electron at the position of different qubit states.
  Over the operation time $\tau$, the effective Hamiltonian
  \begin{equation}
  H=a^{\dagger}_{c_1} a_{c_1}\sum_{l\in p} (\bar{V}_{\text{RF}\ket{1_l}}a^{\dagger}_{l_1} a_{l_1}+\bar{V}_{\text{RF}\ket{0_l}}a^{\dagger}_{l_0} a_{l_0})
  \end{equation}
 accumulates a contrast of $\pi$ phase  on each target atom in $\ket{1}$ qubit state conditioned on the control atom being in state $\ket{1_c}$. 
  Here $a^{(\dagger)}_{l_i}$ annihilates (creates) the Wannier-state of the $l$th target qubit in the plaquette, centred at $l_i$ with $i\in\{0,1\}$ defining the qubit-dependent trap. 
Compensating the background phase \cite{BackPhase} and applying Hadamard to the individual target atoms before and after the Rydberg excitation results in the desired operation of Eq.~\ref{Eq_ParalGate}.

 {\it Gate fidelity:} 
 The main sources of errors in quantifying the C-NOT$^4$ gate's operation are spontaneous emission and population rotation errors. The  errors are averaged over the $2^5$ qubit configurations. 
The average spontaneous emission error from the Rydberg level is given by $E_{sp,r}=\frac{1}{2}\frac{\pi}{U_{\text{RF}} }\Gamma_{\text{r}}$, where $U_{\text{RF}}$ is the qubit-dependent  contrast of $\bar{V}_{\text{RF}}$ and $\Gamma_{\text{r}}$ is the decay rate of the  Rydberg state \cite{Bet09}. In a two-photon excitation scheme of  Fig.~2a, partial population of the intermediate $\ket{6P_{3/2}}$ level results in an extra source of loss. Considering the effective two-photon excitation $\Omega_r=\frac{\Omega_{420}\Omega_{1013}}{2\delta_p}$, using high power lasers \cite{Lev18} facilitates large Rabi frequencies $\Omega_{1013}/2\pi=250$MHz, $\Omega_{420}/2\pi=250$MHz and $\delta_p/2\pi=5$GHz. The corresponding average error would be $E_{se,p}=\frac{\pi}{4\tau_p\delta_p}(q+1/q)=4\times10^{-4}$ \cite{Saf10} with intermediate level lifetime of $\tau_p=113$ns and $q=\Omega_{420}/\Omega_{1013}$.
The control atom's rotation error is due to the unwanted excitation of neighboring accessible Rydberg levels with $\delta_r/2\pi=$17GHz, 21GHz  and 6GHz energy separation in Fig.~1f, 1d and 2 respectively, summing up to an average error of  $1/ 2\frac{\Omega_{r}^2}{4\delta_r^2}$. Finally, non-deterministic excitation of Rydberg atom due to qubit-dependent level-shift caused by Rydberg-Fermi interaction should be overcome by the strong exciting laser $\Omega_{r}$ tuned to the middle of the spectrum leading to average error of $1/2^5\sum \limits_{j=0}^4 \left(\begin{array}{c} 4\\ j \end{array}\right) \frac{  (j-2)^2U_{RF}^2}{\Omega_{r}^2 }$. 
Overall, using the  schemes described in Fig.~1f, 1d, 2d with interactions quantified in Eq. 3, 5, 6 and with the respective Rabi-frequencies $\Omega_r=200,100,40$ MHz results in high fidelity fan-out gate with the average fidelity of F=99.8\%, 99.7\%, 99.6\% at the cryogenic environment of 77K and F= 99.5\%, 99.3\%, 99.2\% without cryogenic environment at 300K. 
The bottleneck in operation fidelity comes from the small lifetime of the Rydberg level. Using Rydberg circular state enhances the interaction to loss ratio by four orders of magnitude, see App.~A.

\subsection{Toffoli gate}  

The {\it Toffoli} gate C$_k$-NOT with $k=4, 6$ could be realized in square and triangular lattices, by placing the control atoms over the plaquette and exciting the central target atom in $\ket{1_t}$ state to the Rydberg level. The Fermi scattering of the Rydberg electron from control atoms forms an interaction-based level-shift on the target atom that depends on the spatial qubit configuration of the entire system. 
Unlike the C-NOT$^k$ gate, Toffoli does not operate with strong laser for deterministic Rydberg excitation. Here a weak transition $\Omega_{r}=\frac{\Omega_{420}\Omega_{1013}}{2\delta_p}\ll V_{\text{RF}}$ would selectively excite the Rydberg atom conditioned on the control atoms to be in  $\ket{0}^{\otimes k}_c$ state. The presence of any $\ket{1_c}$ state localized that control atom inside the Rydberg wave-function of the target atom, shifting the laser out of resonance and blocking the transition. The operation Hamiltonian would be 
\begin{eqnarray}
\label{Eq_Tof}
H_{\text{tof}}=&&(\Omega_1\sigma_{1p}+\Omega_2\sigma_{rp}+h.c.)+\delta_p\sigma_{pp}+\Delta \sigma_{rr}\\ \nonumber
%\Omega_{\text{r}} (a^{\dagger}_{{r_t}}a_{{1_t}}+\text{h.c.})+\Delta a^{\dagger}_{{r_t}}a_{{r_t}}\\ \nonumber
&&+\sigma_{rr} \sum\limits_{l\in p} (\bar{V}_{\text{RF}\ket{1_l}}a^{\dagger}_{l1}a_{l1}+\bar{V}_{\text{RF}\ket{0_l}}a^{\dagger}_{l0}a_{l0}) 
\end{eqnarray}
where $\sigma_{ij}=\ket{i}\bra{j}$ is the transition/projection operator acting on the target atom, $\Omega$ and $\delta_p$, $\Delta$ are the Rabi frequency, and laser detuning from the intermediate and Rydberg levels in a  two-photon excitation. The last term would sum over the control qubits around the central target atoms and apply qubit-dependent Rydberg-Fermi interaction. Adjusting the laser to $\Delta=-k\bar{V}_{\text{RF}\ket{0_l}}$, the $2\pi$  rotation of the target atom would be conditioned on the presence of $\ket{0_c^k1_t}$ state, generates a $\pi$ phase, and apply the desired C$_k$-Z operation. Sandwiching the target atom with Hadamard gates results in the desired Toffoli operation. The Rydberg-Fermi interaction operates the C$_k$-Z  with a single 2$\pi$ pulse addressing the Rydberg level, leaving no Rydberg population unprotected from the laser.
This would eliminate the errors associated with the conventional gate schemes with $\pi$-gap-$\pi$ Rydberg exciting pulses as discussed in \cite{Mal15,Gra19}.

\begin{figure}
\centering 
       \scalebox{0.35}{\includegraphics{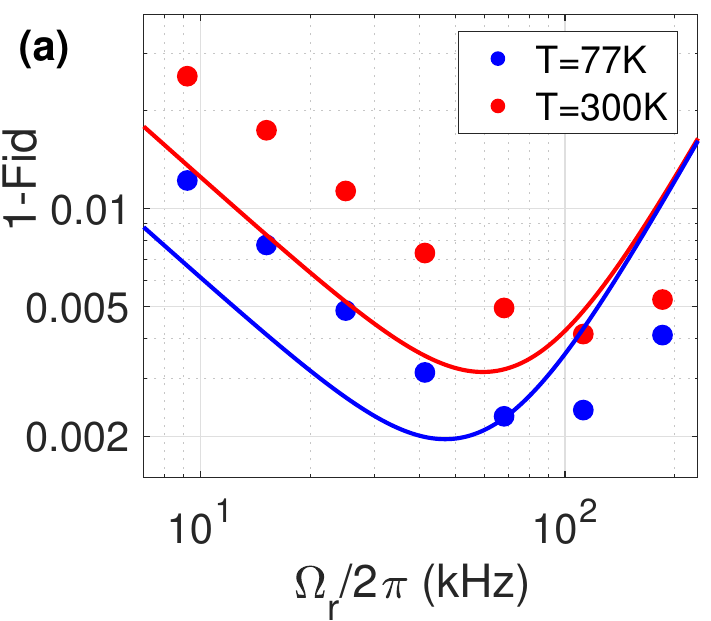}} 
       \scalebox{0.35}{\includegraphics{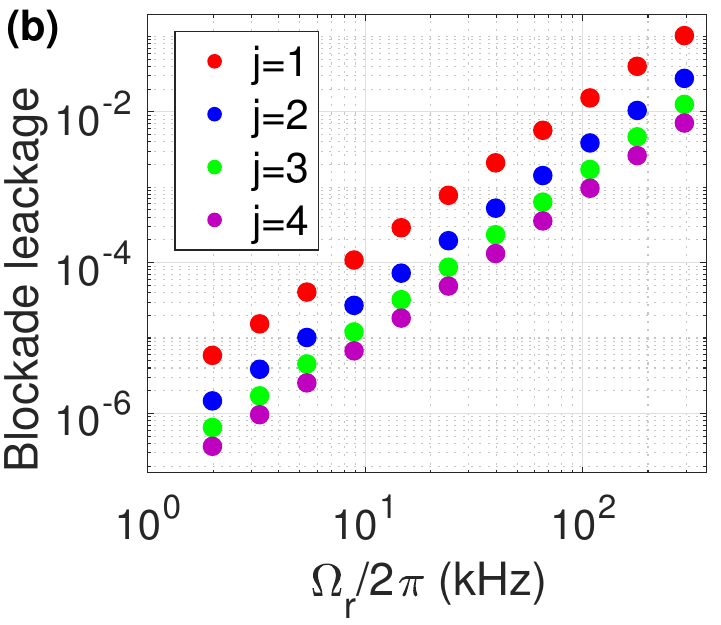}} 
\caption{(a) The infidelity of C$_4$-NOT operation as a function of Rydberg exciting Rabi-frequency for the setup of Fig.~1d with the interaction values quantified in  Eq.~\ref{Eq_64Sup}. The solid-line is based on analytic estimates discussed in the  text for a square laser pulse  while the circles are based on the numerical simulation of Eq.~\ref{Eq_Tof}, calculated for a Gaussian pulse $\Omega_{r}e^{-\frac{(t-T/2)^2}{2\sigma^2}}e^{-\frac{(T/2)^2}{2\sigma^2}}$  with  $\sigma=T/5$ and a pulse duration $T$ given by $\int_0^T\Omega(t)\text{d}t=2\pi$. The fidelity quantified by Eq.~\ref{Eq_FidPhas},  encounters spontaneous emission and rotation errors. The simulation is performed for the case of a cryogenic environment of 77K and at the room temperature of 300K. (b) Blockade leakage of the target atom for different qubit configurations with $j$ plaquette atoms in  $|1_c\rangle$ state. In the numerics, the effective Rabi frequency  $\Omega_r=\frac{\Omega_{420}\Omega_{1013}}{2\delta_p}$  in two-photon excitation is obtained by $\Omega_{420}=\Omega_{1013}$ lasers that are detuned from intermediate level by  $\delta_p/2\pi=5$GHz and the lasers are locked out of phase with the locking bandwidth of 1kHz \cite{Lev18}. The results  are averaged over the frequency profile of the pulse as discussed in App.~D.
}\label{Fig_Operation}
\end{figure}

The main sources of errors in quantifying the C$_k$-NOT gate's operation are spontaneous emission of the atomic levels, and population rotation errors.
The  spontaneous emission from the Rydberg level only occurs in the qubit configuration $\ket{0_c^k1_t}$ where the single target atom gets excited to the Rydberg state. Hence the averaged operation error of $E_{sp,r}=\frac{1}{2^{k+1}}\frac{\pi}{\Omega_{\text{r}}}\Gamma_{r}$ is expected. 
 Off-resonant Rydberg  excitation results in blockade leakage adding up to the average error of $E_{r1}=\frac{1}{2^{k+1}}\sum \limits_{j=1}^k  \left(\begin{array}{c} k\\ j \end{array}\right) \frac{\Omega_{r}^2}{4 j^2 U_{RF}^2}$, where $j$ is the number of control atoms in $\ket{1_c}$ qubit-dependent lattice. 
 %The average infidelity is plotted in Fig.~\ref{Fig_Operation} as a function of laser's Rabi frequency for the  setup discussed in Fig.~1d. %Eq.~\ref{Eq_65P3_2} and (b) Eq.~\ref{Eq_75Sup}. 
 In the two-photon excitation, blockade is sensitive to the locking bandwidth of the lasers, which could be made less than 1kHz \cite{Lev18}. Also in  case of fast operation, the large frequency bandwidth of exciting pulses might affect the blockade at the heart of the scheme. 
Circle signs in Fig.~4 quantify the operation of the gate by simulating  the master equation encountering the spontaneous emission from intermediate and Rydberg levels as well as the de-phasing terms associated with lasers' line-width. Also, the frequency profiles of the laser pulses are encountered in the lasers' detuning and Rabi frequencies as discussed in App.~D. 
 The analytic and numeric simulations of Fig.~\ref{Fig_Operation} suggest high fidelity operations of 99.8\% could be expected in the setup of Fig.~1d. 
 Other avenues in enhancing the fidelity are discussed in App.~A which are based on improving the interaction and lifetime using the resonance scattering in Cs atoms and higher orbital angular momentum quantum numbers.

\subsection{Other sources of error}  
 Rydberg-Fermi scattering could enhance the decoherence rate in the Bose-Einstein condensate (BEC) \cite{Bal13}. This decoherence only occurs at short inter-atomic distances well inside the Rydberg orbital when the attractive Rydberg-Fermi and atom-ion interaction, moves the two interacting atoms to a very small separation of order 2 nm, where the binding energy of the molecules can ionize the Rydberg electron and form a $Rb_2^+$ molecule \cite{Nie15}. Without the mass transport, step-wise decay or ionization of the Rydberg atom is ruled out by the quantization of the Rydberg state, as discussed and experimentally tested in \cite{Bal13}. This is because the small molecular binding energy at the last lobe of the Rydberg wave-function is orders of magnitude smaller than the closest Rydberg levels for the range of principal numbers applied here. The occurrence of ion-pair formation is also highly unlikely in this system \cite{Nie15}. In conclusion, confining the atoms by an optical-lattice at the last lobe of the Rydberg wave-function, prevents the described mass transport and completely closes the Rydberg molecule loss channels. 
Other sources of errors are due to the unwanted entanglement between the motional state and the qubit configurations as well as the imperfections in single site addressing  as discussed in App. D and E.

\section{Discussion}

This article proposes highly controllable multi-qubit operations, based on engineering the electronic cloud of a single Rydberg atom with respect to the atomic lattice.  The new scheme significantly reduces the population of short-lived Rydberg states during the multi-qubit operation compared to other fast conventional dipolar schemes.
Restricting the Rydberg population to a single site eliminates unwanted intra-component interactions, facilitating high-fidelity multi-qubit gates. In a quantitative comparison of the conventional dipolar scheme with the proposed Rydberg-Fermi approach, App.~C shows that the new approach is a significant step in the implementation of single-step multi-qubit operations. 

The proposed Toffoli gate is operating with a continuous 2$\pi$ pulse, leaving no Rydberg population unprotected from the laser. This would eliminate the errors associated with the conventional schemes with $\pi$-gap-$\pi$ Rydberg exciting pulses as discussed in \cite{Mal15,Gra19}.
Another advantage of having a single Rydberg atom is closing the Anomalous broadening decoherence \cite{Zei16,Gol16}. 
Direct implementation of multi-qubit gates in this proposal would reduce the operation steps and hence the accumulative errors. As an example, 
the C$_6$-NOT gate operation with concatenated Rydberg C-NOT gates \cite{Mas03,She09} requires 112 pulses.   Significant contrast obtains in the Rydberg-Fermi scheme operating by three pulses. 

In the outlook, the proposed Rydberg-Fermi interaction paves the way for long-distance entanglement and direct operations on the logical basis \cite{Khaz22Log}.
This would facilitate the investigation of phenomena, and protocols that arise in quantum information over a wide dimension of physical qubit state space while operating on the logical basis that is growing polynomially.

\begin{acknowledgements}
This work was supported by the Austrian Science Fund (FWF) through a START grant under Project No. Y1067-N27 and the SFB BeyondC Project No. F7108-N38, the Hauser-Raspe foundation, and the European Union's Horizon 2020 research and innovation program under grant agreement No. 817482. This material is based upon work supported by the Defense Advanced Research Projects Agency (DARPA) under Contract No. HR001120C0068. Any opinions, findings and conclusions or recommendations expressed in this material are those of the authors and do not necessarily reflect the views of DARPA.
\end{acknowledgements}

\section{Methods}
\subsection{Implementing stabilizer phase-gate using parallelized gate}

The implementation of the stabilizer operator  $B_p=\prod \limits_{i\in p} \sigma_z^{(i)}$ over the plaquette spins applies in three steps i.e.  $B_p=H^{-1}U_gH$ where  $H=\sum\limits_{i\in \{p,c\}} \exp(i \pi/2\sigma^{(i)}_x)$  is Hadamard applying over all the plaquette and control atoms and $U_g$ is the parallelized Ryd-Fermi gate of Eq.~\ref{Eq_ParalGate}. For the control qubit  prepared in $\ket{0}_c$, the gate $B_p$ coherently transfers the control qubit into the state $\ket{1}_c$ ($\ket{0}_c$) for the odd (even) parity of plaquette spin.  The $A_p=\prod \limits_{i\in p} \sigma_x^{(i)}$ stabilizer would be obtained by exclusive application of Hadamard on control atom. 
The desired stabilizer-phase gate, would be implemented by application of a phase shift on the control qubit, sandwiched by applying/reverting the stabilizer operator $B_p$ 
\begin{equation}
U_{\square}(\theta)=e^{i\theta B_p}=B_p^{-1}e^{i\theta \sigma_z^{(c)}} B_p,
\end{equation}
with $\theta$  being optimized between $[0,\pi]$ in QAOA.
At the end, the control atom would be factored out by transferring to the ground state.

\subsection{Data availability}
All data needed to evaluate the conclusions in the article are presented in the article
and/or the Supplementary Figures. Additional data related to this paper may be
requested from the corresponding author.

\subsection{Competing interests}
The authors declare no competing interests.

\subsection{Author contribution}
The project is defined, derived and written by M. Khazali. The project is benefited from the scientific advice of W. Lechner.

\begin{widetext}

\section*{Supplemental material}

\section*{Appendix A: Rydberg cloud engineering in a triangular lattice}

\subsection*{Rydberg states with high orbital angular momentum in rubidium lattice }

\begin{figure*}[h]
\centering 
       \scalebox{0.47}{\includegraphics{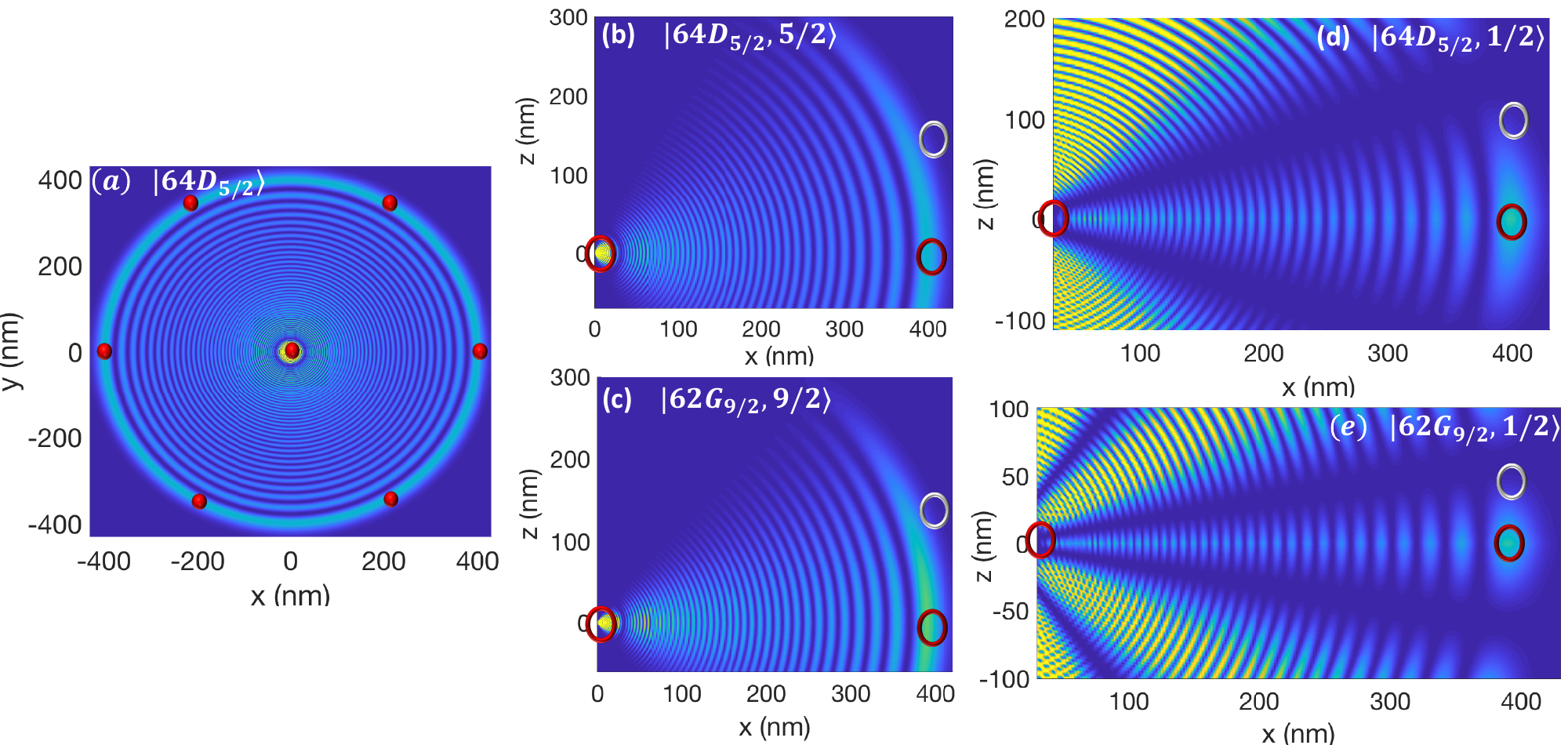}} 
\caption{Designing the Rydberg cloud with respect to the lattice sites depicted by red/white ovals corresponding to the $|0\rangle$/$|1\rangle$ qubit states. (a) Shining the laser perpendicular to the lattice plane (that defines the quantization axis along $z$ ), the symmetry of the wave function in the lattice plane results in a uniform interaction with all plaquette qubits. (b,c) Exciting $|nL_j,m_j=j\rangle$ would confine the cloud close to the lattice plane providing larger interaction. (d,e) Exciting $|nL_j,m_j=1/2\rangle $ increases the number of the angular nodes, allowing significant reduction of the qubit-dependent lattice-shift with the price of reducing the interaction strength. }\label{Fig_Cloud}
\end{figure*}

Going to high orbital angular momentum numbers, the centrifugal force pushes the electron away from the core towards the neighboring ground-state atoms. This could enhance the interaction strength. In the extreme limit, the maximum angular momentum $l=n-1$ in the circular Rydberg state forms an ideal torus wave-function, see below.  

Exciting  $|nL_{j},m_j=j\rangle$ Rydberg state  would exclusively excite the $Y_{L,L}$ spherical harmonic. With the quantization axis being perpendicular to the lattice, the electron wave-function would be confined close to the 2D lattice plane providing a homogenous interaction for all plaquette atoms. The electronic cloud of two Rydberg states $\ket{64D_{5/2},5/2}$ and $\ket{62G_{9/2},9/2}$ are plotted in Fig.~\ref{Fig_Cloud}.
 Please note that  $\ket{nD}$ and $\ket{nG}$  state could be excited via a single photon \cite{Ton04} and double photon quadrupole transitions respectively. Corresponding Rydberg-Fermi interaction and gate fidelities in a triangular lattice are quantified in table I.

\begin{center}
\begin{tabular}{  c c |c c |c c c c |c c}
\hline
\# \qquad \quad  & State   \quad  & \quad $D_z$ \quad & \qquad($\sigma_{\{x,y\}}$,$\sigma_z$) & $ \bar{V}_{\text{RF}\ket{1_l}}$ & $\text{MD}_{V_{\text{RF}\ket{1_l}}}$ &  $ \bar{V}_{\text{RF}\ket{0_l}}$ & $\text{MD}_{V_{\text{RF}\ket{0_l}}}$ & 1-Fid \\
 &  & \quad${\scriptstyle(\text{nm}) }$  \quad &\qquad ${\scriptstyle(\text{nm,nm}) }$ & ${\scriptstyle(\text{MHz}) }$ & ${\scriptstyle(\text{MHz}) }$  &   ${\scriptstyle(\text{MHz}) }$  &  ${\scriptstyle(\text{MHz}) }$  & C$_6$-NOT  \\
 \hline
 1\qquad \quad  & $\ket{64D_{5/2},5/2}$  \quad &  \quad 150  \quad	 & \qquad    (25,30)\qquad   & 1.2 \quad   & \quad 0.13 & 0.17 & 0.09  &  0.002 \\
  \hline
2\qquad \quad  & $\ket{62G_{9/2},9/2}$  \quad &  \quad150  \quad	 &   \qquad   (25,30) \qquad   &   1.7 \quad   & \quad 0.17 & 0.16 & 0.08 & -  \\
  \hline
  3\qquad \quad  & $\ket{64D_{5/2},1/2}$  \quad &  \quad100  \quad	 &   \qquad  (25,30) \qquad   &   0.45 \quad   & \quad 0.047 & 0.04 & 0.015 & 0.003   \\
   \hline
   4\qquad \quad  & $\ket{62G_{9/2},1/2}$  \quad & \quad 45  \quad	 &   \qquad   (20,20) \qquad   &   0.45\quad   & \quad 0.08 & 0.04 & 0.017   & - \\
\hline
\label{TableLattice}
\end{tabular}
\end{center}
Table I: The wavelength of the in-plain optical-Lattice in  \#1,3 is  $\lambda=795$nm  and  in \#2,4 is $\lambda=780$nm  \cite{Aro07}. In calculating the fidelity, spontaneous emission of $nD$ state at 300K environment temperature is considered \cite{Bet09}. 
\\

Further confinement of electron cloud perpendicular to the lattice, allows smaller qubit-dependent lattice-shift $D_z$. This would enhance the Franc-Condone factor and facilitates the qubit rotation on the spin-dependent lattice. Exciting $\ket{nL_j,1/2}$ forms a cloud with $L$ angular nodes. The two examples of $\ket{64D_{5/2},1/2}$ and $\ket{62G_{9/2},1/2}$ are plotted in Fig.~\ref{Fig_Cloud}d,e. These states allow operation in small $D_z$ qubit-dependent lattices with significant overlap of two-qubit Wannier states. The drawback in choosing these types of states is the weak strength of the interaction, see table I.

\subsection*{Realization with Circular states}
The recent advances in fast transition to the Rydberg circular states \cite{Sig17,Car20}, would make them an ideal choice for Rydberg-Fermi gates' application.
The ponderomotive potential of focused Laguerre-Gauss (LG) beams, enables site addressing in exciting circular states  \cite{Car20}. 
Exciting the circular state $\ket{58C}$ of a $^{87}Rb$ atom, the electron would be confined at the position of neighbouring qubits at 176nm, see Fig.~\ref{Fig_Circular}. Considering a plaquette atom with Gaussian ground motional state of FWHM=18nm and qubit dependent lattice shift of $D=50$nm, the interaction would be quantified according to Eq.~\ref{Eq_V01} as
\begin{eqnarray}
\label{Eq_58C}
 &&\bar{V}_{\text{RF}\ket{1_l}}=39 \text{MHz},\quad \text{MD}_{V_{\text{RF}\ket{1_l}}}=4\text{MHz}\\ \nonumber
 &&\bar{V}_{\text{RF}\ket{0_l}}=0.5\text{MHz}, \quad \text{MD}_{V_{\text{RF}\ket{0_l}}}=0.27\text{MHz} \quad \quad
 \end{eqnarray}
 The other advantage of the circular state comes from the minimized overlap of the wave-function with the ionic-core which results in an enhanced lifetime in the order of minutes \cite{Ngu18}. 

  \begin{figure} [h]
\centering 
       \scalebox{.36}{\includegraphics{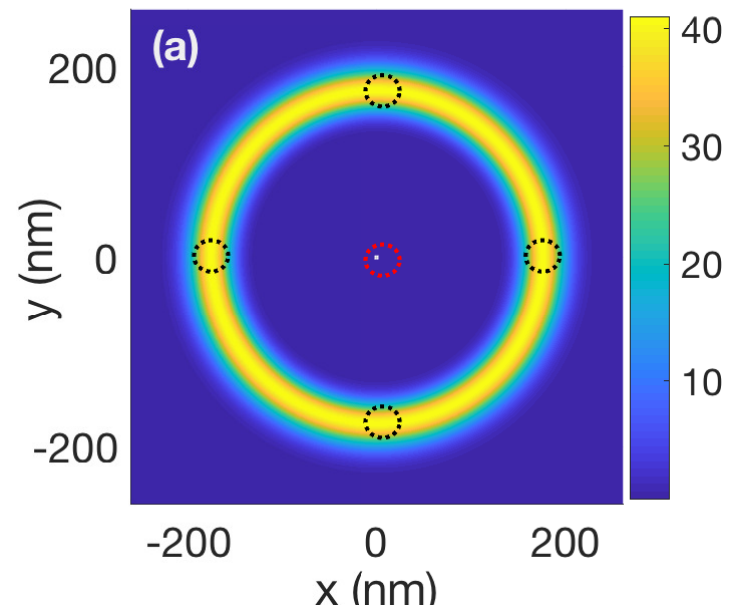}} 
       \scalebox{.36}{\includegraphics{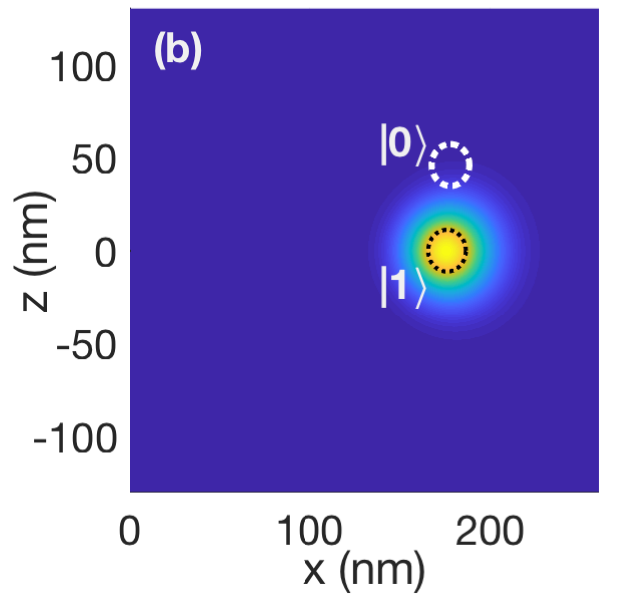}} 
       \scalebox{.32}{\includegraphics{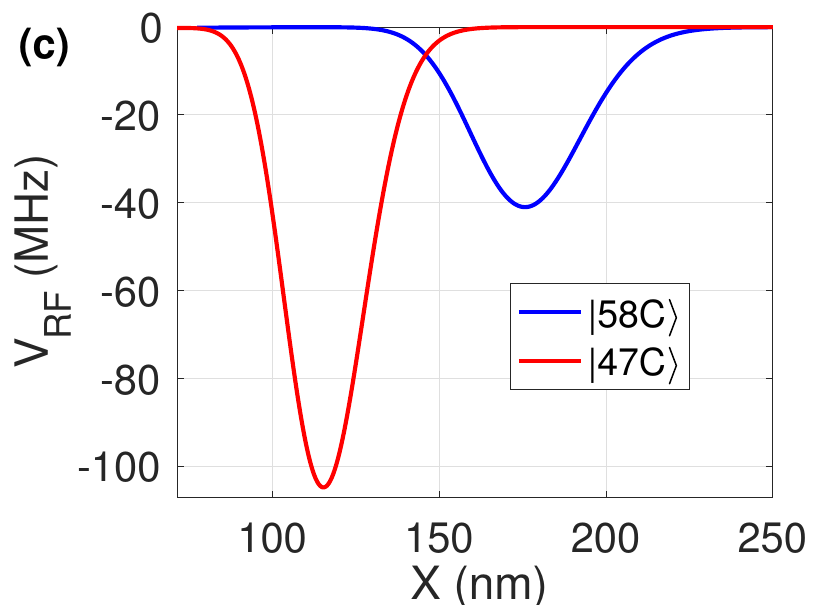}} 
\caption{ Rydberg Fermi interaction with circular states. Exciting $|58C\rangle$,  cross-sections of the interaction are plotted along (a) $xy$ and (b) $xz$. The white and black circles depict the confining area of the $|0\rangle$ and $|1\rangle$ qubit states of the plaquette lattice sites, while the Rb$^{+}$ core is at the origin. (c) The minimal overlap of the interaction profile of $|47C\rangle$ and $|58C\rangle$  could be used for laser switching of the circular Rydberg-Fermi interaction, see the text. }\label{Fig_Circular}
\end{figure}

In an alternative approach one can encode the central atom's $\ket{0_c}$ and $\ket{1_c}$ qubit states in the ground $\ket{g}$ and the circular state $\ket{nC}$. The laser transition between close circular states  $\ket{1_c}=\ket{nC}$ and $\ket{(n+11)C}$ \cite{Coh21}  could be used to turn on the Ryd-Fermi interaction. The radial interaction profile of $\ket{58C}$ and $\ket{47C}$ are plotted in Fig.~\ref{Fig_Circular}c showing  maximum and minimum overlap with a plaquette atom confined at 176nm.
The interaction of $\ket{1_c}$ central qubits would be compensated with global dynamical decoupling (DD) sequences such as WAHUHA \cite{{Wau68}}.

\subsection*{Strong resonance scattering in Cs atoms }

Another approach to enhancing the interaction to loss ratio is harnessing the p-wave near resonance scattering from $^{133}$Cs atoms, available at smaller electron kinetic energies at larger interatomic distances compared to the resonance position in $^{87}$Rb \cite{Boo15}. 
Fig.~\ref{Fig_PEC} plots the potential energy curves (PEC)  of $|46D_{5/2},5/2\rangle+|6S\rangle$ coupled with the neighbouring Rydberg states $|43H+42H+47P+48S\rangle+|6S\rangle$ under Ryd-Fermi interaction 
\begin{equation}
\label{Eq_RydFermi}
V_{\text{RF}}=(2\pi \frac{\tan(\delta^s)}{k(R)}-6\pi\frac{\tan(\delta^p)}{k^3(R)}\stackrel{\leftarrow}{\nabla}_{{\bf r}}.\stackrel{\rightarrow}{\nabla}_{{\bf r}})\delta({\bf r}-{\bf R}).
\end{equation}
Here $\ket{nH}=\sum_{l,m} \ket{n,l,m}$ represents the Hydrogen state encountering semi degenarate orbital angular momentum numbers $2<l<n$.
 The matrix elements in the  manifold of coupled states are given by
\begin{eqnarray}
&&H_{nlm,n'l'm'}({\bf R})=\bra{\psi_{nlm}({\bf R})} V_{\text{RF}}\ket{\psi_{n'l'm'}({\bf R})} \quad \quad \quad \quad \\ \nonumber
&&H_{nlm,nlm}=-\frac{Ry}{n^{*2}}\\ \nonumber
\end{eqnarray}
where $Ry$ is the Rydberg constant of $Cs$ atoms and $n^{*}$ is the effective Rydberg principal number.
Diagonalizing 8000 coupled states, the energy potential is plotted in Fig.~\ref{Fig_PEC}. In a UV optical-lattice with $\lambda=350$nm, the Fermi scattering of Rydberg electron from the neighbouring lattice site would result to about 400MHz level-shift of the Rydberg level ideal for fast quantum operations.

For the p-wave scattering of Rydberg electron from the neighboring ground state atom, the gradient of the Rydberg wave-function $\psi=R_{nl}(r)Y_l^m(\theta,\phi)$ at the position of the neighboring lattice site is required which is
\begin{equation}
 \nabla \psi(r,\theta,\phi)=\begin{bmatrix} 
	\frac{\partial R_{nl}}{\partial r}Y_l^m\\
	\frac{1}{r}R_{nl}\frac{\partial Y_l^m}{\partial \theta} \\
	\frac{1}{r\sin\theta}R_{nl}\frac{\partial Y_l^m}{\partial \phi} \\
	\end{bmatrix}=\begin{bmatrix} 
	\frac{\partial R_{nl}(r)}{\partial r}Y_l^m(\theta,\phi)\\ 
	\frac{1}{r}R_{nl}(r)\frac{1}{2} \sqrt{l^2-m^2}[Y_l^{m+1}(\theta,\phi)e^{-\text{i}\phi}- (l+m+1)Y_l^{m-1}(\theta,\phi)e^{\text{i}\phi}] \\ 
	\text{i}m\frac{\psi(r,\theta,\phi)}{r\sin(\theta)}\\ 
\end{bmatrix} \quad\quad\quad\quad%\quad\,\quad~\,\,\,\,\,\,\,\,\,\,\,\,\,\,\,\,\,\,\,\,\,\,\,\,
	\end{equation}
%\begin{equation}
%\begin{bmatrix} \nonumber
%	\frac{\partial R_{nl}(r)}{\partial r}Y_l^m(\theta,\phi)\\ \nonumber
%	\frac{1}{r}R_{nl}(r)\frac{1}{2} \sqrt{l^2-m^2}[Y_l^{m+1}(\theta,\phi)e^{-\text{i}\phi}- (l+m+1)Y_l^{m-1}(\theta,\phi)e^{\text{i}\phi}] \\ \nonumber
%	\text{i}m\frac{\psi(r,\theta,\phi)}{r\sin(\theta)}\\ \nonumber
%\end{bmatrix} \nonumber
%\end{equation}
in the spherical coordinate. The radial wave-function and its derivative are calculated numerically using the Numerov technique \cite{Gal94}.

  \begin{figure}
\centering 
       \scalebox{0.54}{\includegraphics{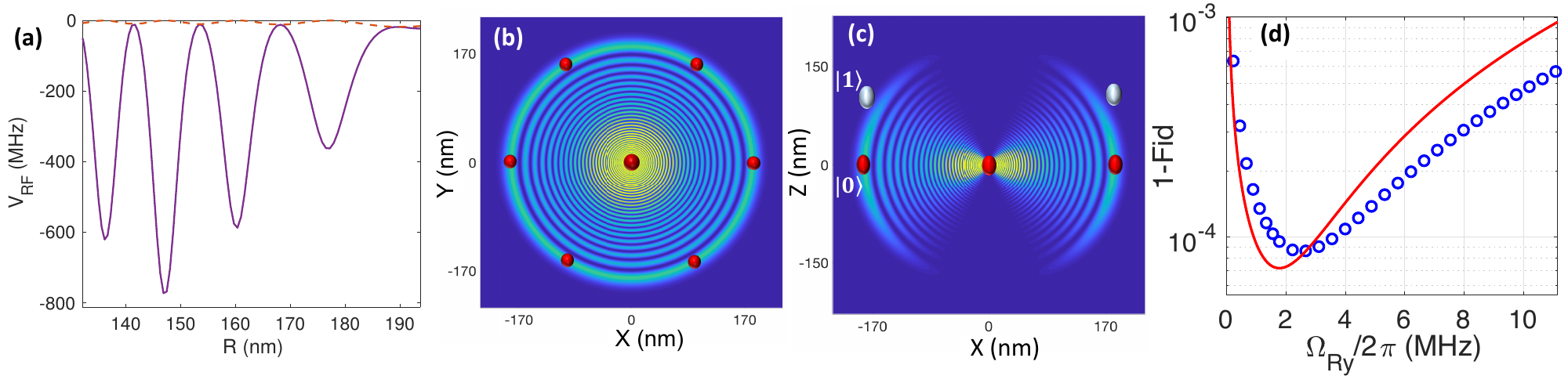}} 
\caption{ (a) PEC with S- and P-wave scattering in Cs atoms.  The coupling of the Rydberg state $|46D_{5/2},5/2\rangle$ with the neighbouring states  $|43H+42H+47P+48S\rangle$ is considered under Eq.~\ref{Eq_RydFermi}. Interaction strength is plotted across the radial direction from the Cs$^{+}$ core with $\theta=\pi/2$ in spherical coordinate. The solid line and dashed line show the total and exclusive s-wave scattering respectively.
(b,c) The  cross-sections of $|46D_{5/2},5/2\rangle$ Rydberg wave-function along (b) $xy$ and (c) $xz$ are plotted. The red and white circles depict the confining area of the $|0\rangle$ and $|1\rangle$ qubit states of a plaquette lattice site, while the Cs$^{+}$ core is at the origin. (d) The infidelity of C$_6$-NOT operation as a function of Rydberg exciting Rabi-frequency. The solid line is based on analytic model for a square laser pulse  while the circles are based on the numerical simulation of Eq.~\ref{Eq_Tof}, obtained for a Gaussian pulse $\Omega_{r}e^{-\frac{(t-T/2)^2}{2\sigma^2}}e^{-\frac{(T/2)^2}{2\sigma^2}}$  with  $\sigma=T/5$ and a pulse duration $T$ given by $\int_0^T\Omega(t)\text{d}t=2\pi$. The fidelity quantified by Eq.~\ref{Eq_FidPhas},  encounters spontaneous emission and rotation errors.
}\label{Fig_PEC}
\end{figure}

In a triangular lattice of Fig.~\ref{Fig_PEC}b, the in-plane $x-y$ Cs trap is formed by 350nm UV laser dressing the ground state to $\ket{10P}$ state. %Applying strong trap potential, of $U_{\text{trap}}/2\pi=20$MHz 
Considering the Gaussian ground motional state with a half-width at 1/e maximum  of $\sigma_{x,y}=8.7$nm,  the atom would accommodate within a single lobe of $V_{\text{RF}}$, see Fig.~\ref{Fig_PEC}b,c. Alternative trapping schemes are discussed in \cite{Alternative}.  
 For spin-dependent trap perpendicular to the plane (along the $z$) the $\lambda=870$nm laser could be used for dressing $\ket{6S}$ to $\ket{6P}$, with $U/2\pi=2$MHz, $\sigma_z=20$nm, $D_z=100$nm.
Exciting the target atom to  $|46D_{5/2},5/2\rangle$, the  Rydberg-Fermi interaction averaged over the l$^{th}$ plaquette atom's wave-function in the ground motional state would be
\begin{eqnarray}
\label{Eq_46D}
 &&\bar{V}_{\text{RF}\ket{1_l}}=365\text{MHz},\quad \text{MD}_{V_{\text{RF}\ket{1_l}}}=0.07\text{MHz}\\ \nonumber
 &&\bar{V}_{\text{RF}\ket{0_l}}=2\text{MHz}, \quad \text{MD}_{V_{\text{RF}\ket{0_l}}}=0.06\text{MHz} \quad \quad
 \end{eqnarray}
 see Eq.~\ref{Eq_V01} for the definitions.  This contrast of spin-dependent level-shift with narrow lines allows fast selective laser excitation of central atom conditioned on the  plaquette qubits' configurations. This would result in high fidelity operation of C$_6$-NOT gate as depicted in Fig.~\ref{Fig_PEC}d. The fidelity is quantified along the same lines described in Fig.~\ref{Fig_Operation}.

 Single site addressing requires ultra strong focusing with large NA microscope \cite{Bak09,Bev01} , see App.~E. Alternatively, the sub-wavelength localized population rotation via semi interferometer techniques \cite{Aga06,Cho07} or dual-species atomic lattice \cite{Huf22,Khaz22Log,She21,Sin21,LHZ15} with central $^{87}$Rb and plaquette $^{133}$Cs atoms could be used to improve single site addressing.

 %This proposal requires further compact lattice structure and in-site confinement. Using an immersion microscope with NA=1.3 \cite{Bev01}   allows focusing the 298nm UV laser to the half-width at half-maximum of 70nm. Hence the neighboring sites at about 200nm would experience four orders of magnitude smaller light intensity.   
 %Alternatively, the sub-wavelength localized population rotation via semi interferometer techniques \cite{Aga06,Cho07} or dual-species atomic lattice \cite{She21,Sin21,LHZ15} with central $^{87}$Rb and plaquette $^{133}$Cs atoms could be used to improve single site addressing. 

\section*{Appendix B: Qubit-Rotation in the spin-dependent lattice }

{\it Qubit-rotation in the spin-dependent lattice --}
A spin-dependent lattice provides dual spin/spatial encoding of the qubit. 
A Raman transition coherently transfers an atom from one internal state to the other, thereby causing hopping between the two Wannier-functions \cite{Jak98,Jak03,Maz10}, see Fig.~\ref{Fig_SpinRotation}d. 
The polarizability of the qubit states $\ket{0}$, $\ket{1}$ and the intermediate electronic-level $\ket{5P_{1/2},1/2}$ in the optical lattice are  given by different light polarization elements, see Fig.~\ref{Fig_SpinRotation}a,b. Hence by tuning the polarization angle $\theta$ between counter-propagating linearly polarized lights it is possible to confine the intermediate state between the two qubit-dependent lattices, see Fig.~\ref{Fig_SpinRotation}c,d.

Here the Raman assisted transition of a trapped neutral atom between the two spin-dependent lattices centred at $l_0$ and $l_1$ is analyzed. 
 A Raman transition between atom's internal states $\ket{0}$ and $\ket{1}$ makes the atom experience a different trap shifted by $D$ where the initial and final vibrational wave-functions have overlap. This scheme resembles the Franck-Condon principle in molecular physics. Under the Born-Oppenheimer approximation, the electronic and the nuclear motions are separated and hence the effective wave-function would be presented as a product of the electronic wave-function and the vibrational wave-function 
\begin{equation}
\psi_{l_i}({\bf R},{\bf r})=w({\bf R}-{\bf l}_i)\psi_{e,i}({\bf R},{\bf r})
\end{equation}
where ${\bf r}$ and ${\bf R}$ are addressing the electronic and centre of mass positions. The Wannier function of the $l$th site in the $i\in\{0,1,p\}$ spin dependent lattices is given by $w({\bf R}-{\bf l}_i)$.  
The electric dipole transition from a state A to an excited state B is given by
\begin{equation}
\langle \psi_{A}|({\bf R}+{\bf r})|\psi_B\rangle=\langle w({\bf R}-{\bf l}_A)|w({\bf R}-{\bf l}_B)\rangle  \langle \psi_{e,A}|{\bf r} |\psi_{e,B}\rangle.
\end{equation}
Here we have the orthogonality of the electronic eigenstates but the vibrational states are belonging to different traps and do not need to be orthogonal. Also using the  Condon approximation, the dipole transition of electronic states is assumed independent of nuclear coordinates.
In conclusion, the dipole transition would be modified by the overlap of the wave functions i.e.  Franck-Condon factor $f=  \langle w({\bf R}-{\bf l}_A)|w({\bf R}-{\bf l}_B)\rangle$.

\begin{figure}
\centering 
     \scalebox{0.52}{\includegraphics{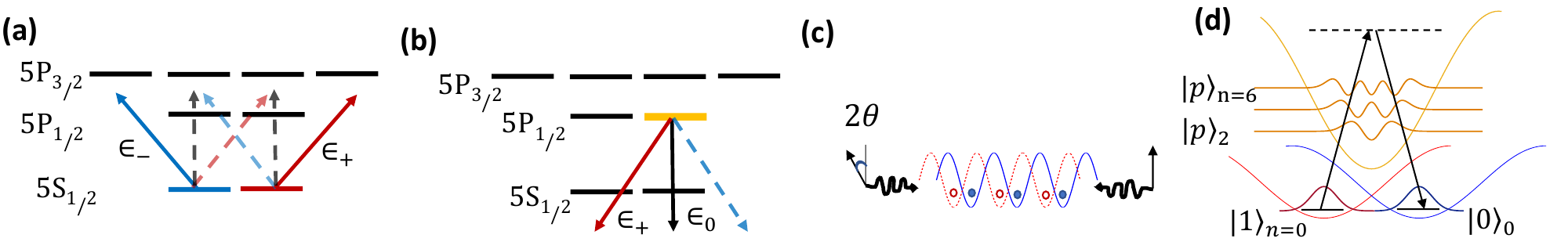}} 
\caption{Qubit rotation in the spin-dependent lattice. (a)   In Rb (Cs), tuning the trapping laser between  $5P_{3/2}$ ($6P_{3/2}$) and $5P_{1/2}$  ($6P_{1/2}$), the polarizability of qubit states $|0\rangle$ and $|1\rangle$ are given by distinguished circularly polarized lights $\varepsilon_{-}$ and $\varepsilon_{+}$ respectively. (b) The same fields, trap the $|p\rangle=|5(6)P_{1/2},1/2\rangle$ electronic state with $\varepsilon_{0}$ and $\varepsilon_{+}$ elements.  (c)  The relative polarization of $2\theta$ between counter propagating linearly polarized lights could be tuned to trapp the intermediate state between the qubit states (d).  The Raman transition in a dual spin/spatial encoded qubit would be modified by the Frank-Condon factor. }\label{Fig_SpinRotation}
\end{figure} 

{\it Effective qubit-rotation rate:} The Hamiltonian  $H=H_0+H_d$ consists of  the energy level of electronic states $H_0$, and the dipole transitions $H_d$. The Hamiltonian of the system in the rotating wave approximation and in the basis $\{ \ket{0}_0,\ket{1}_0,\ket{p}_0,\ket{p}_1,\ket{p}_2,...,\ket{p}_n \}$ would be
\begin{equation}
 \tilde{H}=\begin{bmatrix}
   0 &0&f_{0p_0} \Omega_0/2 & f_{0p_1} \Omega_0/2&\dots &f_{0p_n} \Omega_0/2  \\
   0 &\delta& f_{1p_0} \Omega_1/2& f_{1p_1} \Omega_1/2& \dots &f_{1p_n} \Omega_1/2 \\
   f_{0p_0}\Omega_0/2 &   f_{1p_0}\Omega_1/2 & \Delta-1/2\omega_{tr}&0&\dots&0\\ 
   f_{0p_1}\Omega_0/2 &   f_{1p_1}\Omega_1/2 & 0&\Delta-3/2\omega_{tr}&\dots&0\\ 
   \vdots &   \vdots & \vdots & \vdots & \ddots & \vdots \\ 
      f_{0p_n}\Omega_0/2 &   f_{1p_n}\Omega_1/2 & 0&0& \dots&\Delta-(n+1/2)\omega_{tr}\\ 
   \end{bmatrix}
\end{equation}
where $\Delta=\omega_{L0}-(\omega_p-\omega_0)$, $\delta=(\omega_{L0}-\omega_{L1})-(\omega_{1}-\omega_{0})$. 
The Frank-Condon factors $f_{in}=\int w_i^{*}({\bf x}-{\bf l_{i}}) w_{p_n}({\bf x}-{\bf l_p})\text{d}{\bf x}$ quantify the overlapping of the qubit states $i=\ket{0,1}_{n=0}$ and intermediate $\ket{p}_{n}$ state's  Wannier function $w_{p_n}$  in the  $n$th motional state.
The Schr\"odinger equation in the interaction picture could be written in terms of coupled equations:
\begin{eqnarray}
&&\frac{\text{d}C_0}{\text{d}t}=\text{i}\sum\limits_{j=0}^n \frac{f_{0j} \Omega_0}{2} C_{p_j}\\
&&\frac{\text{d}C_1}{\text{d}t}=\text{i}\sum\limits_{j=0}^n \frac{f_{1j}\Omega_1}{2} C_{p_j}\\
&&\frac{\text{d}C_{p_j}}{\text{d}t}=\text{i}(\frac{f_{0j}\Omega_0}{2} C_{0}+  \frac{f_{1j}\Omega_1}{2}  C_{1}+[\Delta-(j+1/2)\omega_{tr}]C_{p_j})\\
&&\end{eqnarray}
Under the condition that $\Delta$ is the dominant term the amplitudes $C_{p_j}$  would oscillate much faster than $C_{0}$ and $C_{1}$. Hence we can apply adiabatic elimination by substituting 
\begin{equation}
C_{p_j}=-\frac{f_{0j}\Omega_0 C_{0}+  f_{1j}\Omega_1  C_{1}}{2(\Delta-(j+1/2)\omega_{tr})} 
\end{equation}
into equations 10, 11.  Hence the system would be represented by an effective two-level system:
\begin{eqnarray}
&&\frac{\text{d}C_0}{\text{d}t}=-\text{i}\sum\limits_{j=0}^n\frac{f_{0j}^2\Omega_0^2 \, C_{0} + f_{0j}\Omega_0f_{1j}\Omega_1 \, C_1}{4(\Delta-(n+1/2)\omega_{tr})} \\
&&\frac{\text{d}C_1}{\text{d}t}=-\text{i}\sum\limits_{j=0}^n\frac{f_{0j}\Omega_0 f_{1j}\Omega_1 \, C_{0} + f_{1j}^2 \Omega_1^2 \, C_1}{4(\Delta-(n+1/2)\omega_{tr})} +\text{i}\delta C_1
\end{eqnarray}
Assuming that $\Delta\gg U_{tr}$  the effective Rabi frequency in the two-level system would be 
\begin{equation}
\tilde{\Omega}=\frac{\Omega_0\Omega_1}{4\Delta} \sum\limits_{j=0}^n f_{0j}f_{j1}
\end{equation}
and the effective detuning of the two-level system  would be 
\begin{equation}
\delta_{\text{eff}}=\delta-\sum\limits_{j=0}^n \frac{f_{1j}^2\Omega_1^2-f_{0j}^2\Omega_0^2}{4\Delta}
\end{equation}
\end{widetext}
The qubit-rotation is performed in the regime of $\tilde{\Omega}\ll \omega_{\text{tr}}$ to avoid exciting the motional Bloch bands.

 \section*{Appendix~C: Comparing Rydberg-dipolar and Rydberg-Fermi gates }
 \label{Sec_UnwantedPhase}
 
 \subsection*{Rydberg decay} 

The main bottleneck in the Rydberg quantum computation/simulation is the short lifetime of Rydberg levels. 
The Rydberg dipolar multi-qubit Toffoli C$_k$-NOT gate could be realized by two approaches as explained in \cite{Ise11}.  In the fast scheme, all control atoms in $\ket{0}$ qubit state would get excited to the Rydberg level with simultaneous pulses. In that case, the control atoms would be out of the intra-component blockade and the average population of Rydberg levels in control atoms over the $2k+1$ qubit configurations would scale by k leading to an average Rydberg decay error of $k\Gamma(\pi/2\Omega_c +3\pi/2\Omega_t )$ see Eq.~5 in  \cite{Ise11}. In the other scheme operating with sequential excitation steps, the Rydberg population is limited to one due to the global Blockade effect but the $2k+3$ sequence of excitation would result in a long operation time leading to the same scaling of average Rydberg decay error of control atoms $k\Gamma\pi/2\Omega$ \cite{Ise11}. This is in contrast to the Rydberg-Fermi scheme where the only qubit configuration with a single Rydberg population is $\ket{0_c^k1_t}$ qubit state. In the other $2^{k+1}-1$ qubit configurations, the target laser would be out of resonance and no population would be excited. Hence the Rydberg decay error averaged over all qubit configurations would be  $\frac{1}{2^{k+1}}\pi\Gamma/2\Omega$. Please note that the scheme could operate for $k=6$ in a triangular lattice and $k>6$ by increasing the principal number. This results in a significant difference compared to the dipolar scheme.

 \subsection*{Phase errors due to intra-component interaction} 

 One of the advantages of implementing multi-qubit operations with the single step Rydberg-Fermi scheme comparing to the fast Ryd-dipolar counterpart is in the absence of intra-component interaction. To quantify the effects of this unwanted phase on gate performance, the  phase sensitive form of fidelity is used 
\begin{equation}
\label{Eq_FidPhas}
\text{Fid}=\text{Tr}(|M+MM^{\dagger}|)/2n,
\end{equation}
where $M=U_{id}^{\dagger}U_{gate}$, with $U_{id}$ and $U_{gate}$ representing the ideal and realistic gate operations. Dimension of the qubit configurations in C$_k$-NOT or C-NOT$^k$  is given by $n=2^{k+1}$. 

  \begin{figure}[h]
  \centering 
       \scalebox{.33}{\includegraphics{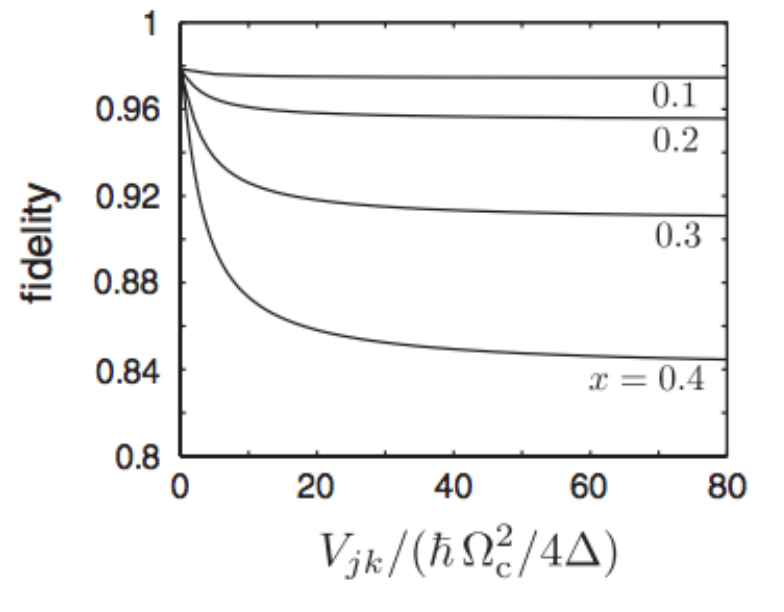}} 
\centering 
       \scalebox{.29}{\includegraphics{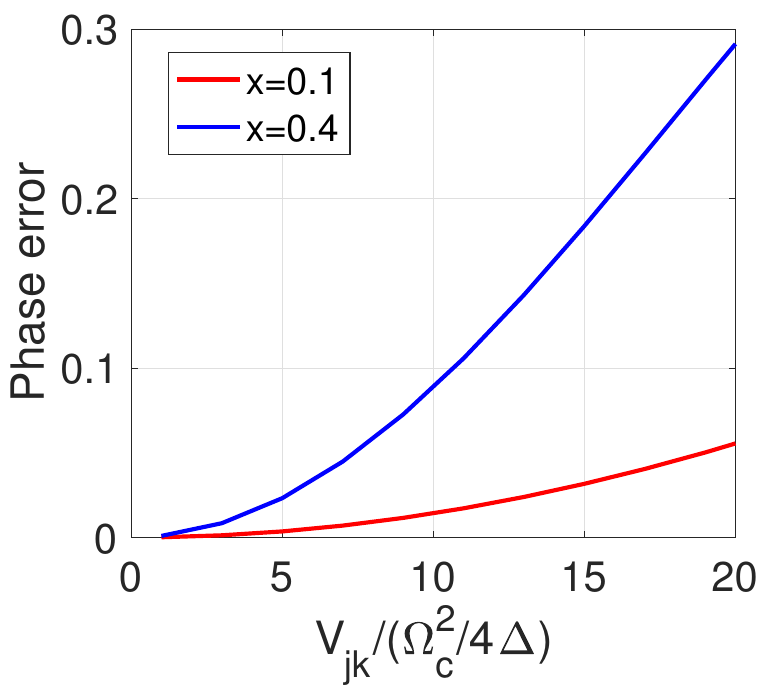}} 
\caption{ Effects of unwanted intra-component interaction on the parallelized gate operation. (a) Neglecting the phase error, Ref. \cite{Mul09} quantifies gate operations in making three-qubit GHZ state for different $x=\sqrt{2}\Omega_p/\Omega_c$. The figure suggests high fidelity in the presence of strong intra-component interaction among target atoms $V_{jk}$.  Using this gate for implementing the stabilizer operator \cite{Wei10} makes the operation phase-sensitive. (b)
Applying the phase-dependent definition of fidelity in Eq.~\ref{Eq_FidPhas}, this dipolar Rydberg gate  \cite{Mul09} is significantly sensitive to $V_{jk}$. The proposed Ryd-Fermi scheme in this article is an alternative with no cross-talk among target atoms.}\label{Fig_PhaseErrorMuller}
\end{figure}

Evaluating the fidelity of Rydberg-dipolar {\it Toffoli gate} C$_4$-NOT proposed in \cite{Ise11}, with the phase dependent definition of fidelity in Eq.~\ref{Eq_FidPhas} reveal the effects of unwanted phase.
In \cite{Ise11}, control and target atoms are getting excited to $|60S\rangle$ and $|60P\rangle$, with optimum laser couplings of $\Omega_c/2\pi=180$MHz, $\Omega_t/2\pi=0.8$MHz and lattice separation of 4$\mu$m. Simulating the gate operation under the Schr\"odinger equation, encountering spontaneous emission and population rotation errors in addition to the infidelity encountered by unwanted phases leads to the average infidelity of 5\% quantified by Eq.~\ref{Eq_FidPhas}.
 Importantly, large phase-dependent infidelities would occur for specific qubit configurations with large  Rydberg population e.g. $\ket{1111}_c\ket{1(0)}_t$ experience 52\% (12\%) infidelity. This is while the phase independent conventional definition leads to 2\% average infidelity.  
 In the proposed Ryd-Fermi approach the absence of intra-component interaction, reduced rotation errors at the lower principal numbers, and reduced Rydberg population results in high-fidelity operations that surpass the conventional counterparts \cite{Kha20,Ise11} while the scheme allows orders of magnitude compression of the atomic quantum processor.   Notably, the assigned fidelity is below the 1.4\% infidelity threshold for surface error correction codes.

The implementation of Rydberg-dipolar {\it parallelized gate} \cite{Mul09} is also sensitive to the intra-component interaction. Using the parameters of Ref. \cite{Mul09}, in the absence of the Rydberg control atom, each of the target atoms would follow the dark state $\ket{D}=(\Omega_c\ket{A}-\Omega_p\ket{R})/N$ with the Rydberg population of $P_R=(\frac{\Omega_p}{\Omega_c})^2$ on each target atom. For the abbreviation, the readers are referred  to \cite{Mul09} for the scheme explanation and parameter definitions.  
With phase independent definition of fidelity, Ref.~\cite{Mul09}  predicts the gate operation to be insensitive to strong intra-component interaction $V_{jk}$, see Fig.~\ref{Fig_PhaseErrorMuller}a. However, using the parallelized gate for the implementation of the stabilizer operator \cite{Wei10} makes the operation phase-sensitive. Applying phase-dependent definition of fidelity Eq.~\ref{Eq_FidPhas}, the dipolar Rydberg gate is significantly sensitive to $V_{jk}$ as shown in Fig.~\ref{Fig_PhaseErrorMuller}b. This comparison shows the value of Rydberg-fermi gates in the implementation of quantum simulation and optimization algorithms.

 \section*{Appendix~D: Adiabaticity and pulse duration }
 \label{Sec_Adiabaticity}

In the parallelized gate, the Rydberg-Fermi potential modifies the optical trapping experienced by the plaquette atoms in $\ket{1_t}$ Wannier states. It is important to apply the changes adiabatically to avoid an unwanted entanglement between the computational and the motional states.   Over the  Rydberg excitation of the control atom,  target atoms in  $\ket{1_t}$ Wannier state would experience trap evolution $U_{trap}=U_{op}+P_{r_c}(t)V_{\text{RF}}$ where $U_{op}$ is the optical trapp, $P_{r_c}$ is the Rydberg population of the central atom. As long as the dynamic is adiabatic, i.e. $\dot{\omega}_{trap}\ll \omega^2_{trap}$ \cite{Wit20}, the Wannier states of the $\ket{1_t}$  can  adapt continuously and stays close to the instantaneous ground motional state. Operating the C-NOT$^k$ with the setup associated with Fig.~2/Eq.~6, a linear change of Rabi frequency from $\Omega_r=30$MHz to $\Omega_r=45$MHz would preserve the ground motional state. 
Unlike the parallelized gate in Toffoli, the presence of a plaquette atom in the Rydberg wave-function blocks the Rydberg excitation. Hence no bound state would be formed and the above adiabaticity discussion does not limit the Toffoli scheme.
%Finally, while preserving the ground motional state in the optical lattice platform helps in tuning the Rydberg-Fermi interaction, applying the dark-state optical-lattice \cite{Yav09} eliminates the concerns in preserving the ground motional states. 
Stronger confinement of atoms in dark-state optical-lattices \cite{Yav09} or twist-optical-lattices \cite{Kha23} with sub-nanometer trap width, allows faster adiabatic operations.

The other concern about pulse duration is related to the pulse bandwidth. While fast operation makes short pulses desirable, short pulses would be wide in bandwidth and might excite the neighboring Rydberg levels  \cite{Tak16} in fan-out or disturb the blockade in Toffoli gate.
For the chosen Rydberg levels $\ket{65P_{3/2}}$, $\ket{64D}$ and $\ket{75D}$ in Fig.~1 and 2, the level spacing to the next dipole accessible Rydberg level would be 17GHz, 21GHz and 6 GHz respectively.
% While the Toffoli gate operates with long pulses in the order of $\tau \gtrsim$40$\mu$s the
  The fan-out gate operates with stronger Rabi frequencies compared to Toffoli and in principle could operate with shorter pulses $\tau \gtrsim$140ns. Corresponding pulse bandwidths would be at least three orders of magnitude smaller than the level spacing in the above-mentioned cases.
%Also in Toffoli, the pulse bandwidth corresponding to the optimum Rabi frequency of Fig.~4 is 40 to 160 times less than the  interaction-induced level-shift ensuring the perseverance of blockade. $\clubsuit$

In the fast operation of the Toffoli gate, the laser pulse bandwidth might be comparable with the interaction-induced level-shift suppressing the blockade at the heart of the scheme.
Circle signs in Fig.~4 quantify the gate's operation by simulating  the master equation encountering  the pulse bandwidth, the  spontaneous emissions in two-photon excitation, and the de-phasing terms associated with laser line-widths as described below. The effective Rabi frequency in two-photon excitation $\Omega_r=\frac{\Omega_{420}\Omega_{1013}}{2\delta_p}$   is obtained by $\Omega_{420}=\Omega_{1013}$ lasers that are detuned from intermediate $\ket{p}$ level by  $\delta_p/2\pi=5$GHz.
 The Gaussian pulses $\Omega_{r}e^{-\frac{(t-T/2)^2}{2\sigma^2}}e^{-\frac{(T/2)^2}{2\sigma^2}}$ are considered in the numerics with $\sigma=T/5$ and a pulse duration $T$ given by $\int_0^T\Omega(t)\text{d}t=2\pi$. 
 Over the two-photon excitation, the Fourier transform of the two laser pulses would be $\Omega_1(\omega_{l1})=\Omega_{420}e^{-\frac{\omega_{l1}-\omega_{1c}}{4w^2}}$ and $\Omega_2(\omega_{l2})=\Omega_{1013}e^{-\frac{\omega_{l2}-\omega_{2c}}{4w^2}}$ with $\omega_{c}$ indicating the central frequency of the pulse and the pulse bandwidth is presented by $w=1/\sigma$. 
   
 The driving Hamiltonian of Eq.~\ref{Eq_Tof}  is a function of the pulse frequency elements $\omega_{l2}$ and $\omega_{l1}$ both in Rabi frequencies and detunings. The Fourier transform of the two laser pulses would be $\Omega_1(\omega_{l1})=\Omega_{420}e^{-\frac{\omega_{l1}-\omega_{1c}}{4w^2}}$ and $\Omega_2(\omega_{l2})=\Omega_{1013}e^{-\frac{\omega_{l2}-\omega_{2c}}{4w^2}}$ with $\omega_{c}$ indicating the central frequency of the pulse and the pulse bandwidth is presented by $w=1/\sigma$. 
 The laser detunings from the intermediate and Rydberg levels  $\delta_p=\omega_{l1}-\omega_{1p}$ and $\Delta=\omega_{l1}+\omega_{l2}-\omega_{1r}$ are also a function of the pulse frequency elements.
 In Fig. 4a,b the master equation (Eq.~\ref{Master}) is simulated for distinct pulse frequencies $\omega_{l1}$, $\omega_{l2}$ and the final results are averaged over the Gaussian frequency profile of the two laser pulses.

In the operation, the target atom is subject to de-phasing and decay terms that are encountered by the master equation
\begin{eqnarray}\label{Master}
\partial_t \hat{\rho}=&&-\text{i}[\hat{H},\hat{\rho}] + \mathcal{L}(\hat{\rho})
\end{eqnarray}
where the Liouvillian term $\mathcal{L}(\rho)=\sum\limits_\beta \mathcal{D}(c_{\beta})\rho$ with $\mathcal{D}(c)\rho=c\rho c^{\dagger}-1/2 (c^{\dagger}c \rho+\rho c^{\dagger}c )$ in the Lindblad form governs the dissipative time evolution. 
%Lindblad terms encounter spontaneous emission from the Rydberg and  intermediate levels to the qubit and non qubit states. To obtain a pessimistic limit of fidelity we consider the entire decoherence as a loss of population to other electronic states $c_{p}=\sqrt{\gamma_p}|o\rangle \langle p|$ and  $c_{r}=\sqrt{\gamma_r }|o\rangle \langle r|$.
Lindblad terms encounter spontaneous emission from the intermediate level to the qubit states $c_{1p}=\sqrt{\gamma_p/2}|1\rangle \langle p|$, $c_{0p}=\sqrt{\gamma_p/2}|0\rangle \langle p|$  as well as the loss of population from Rydberg state to other electronic states $c_{or}=\sqrt{\gamma_r }|o\rangle \langle r|$. 
Furthermore, the dephasing terms associated by the lasers' linewidth are included as  $c_{11}=\sqrt{\Gamma_{11}}|1\rangle \langle 1|$, $c_{pp}=\sqrt{\Gamma_{pp}}|p\rangle \langle p|$, $c_{rr}=\sqrt{\Gamma_{rr}}|r\rangle \langle r|$,
  where $\Gamma_{11}=(\gamma_{\text{Lock}}+\gamma_{l1}-\gamma_{l2})/2$, $\Gamma_{pp}=(-\gamma_{\text{Lock}}+\gamma_{l1}+\gamma_{l2})/2$, $\Gamma_{rr}=(\gamma_{\text{Lock}}-\gamma_{l1}+\gamma_{l2})/2$ \cite{Khaz21} with $\gamma_{l1}$ and $\gamma_{l2}$ being the line-widths of $\Omega_{1}$ and $\Omega_{2}$ lasers (corresponding to 420nm and 1013nm lasers respectively). %By locking the lasers, the fidelity would be preserved over a wider range of laser noise even above EIT bandwidth \cite{}.
  The coherence in two-photon excitation would be sensitive to the linewidth of the Lock $\gamma_{\text{Lock}}$  when the lasers are locked out of phase  \cite{Khaz21}, which could be suppressed to less than 1kHz \cite{Lev19}.

 \section*{Appendix~E: Single site addressing}
 
 \subsection{Applying local light-shift}
 The Laser cross-talk and misalignment could affect the accuracy of gate operation in compact lattices. The population that does not return to the qubit basis over the Rydberg excitation would be considered as loss. Inspired by \cite{Wei11}, single-site addressing could be performed by applying site-selective differential light-shift to the $\ket{1}\bra{r}$ transition. Focusing the 788nm auxiliary laser to the targeted site, only the desired atom would get  in-resonance with the Rydberg exciting laser.

\begin{figure}[h]
  \centering 
         \scalebox{.35}{\includegraphics{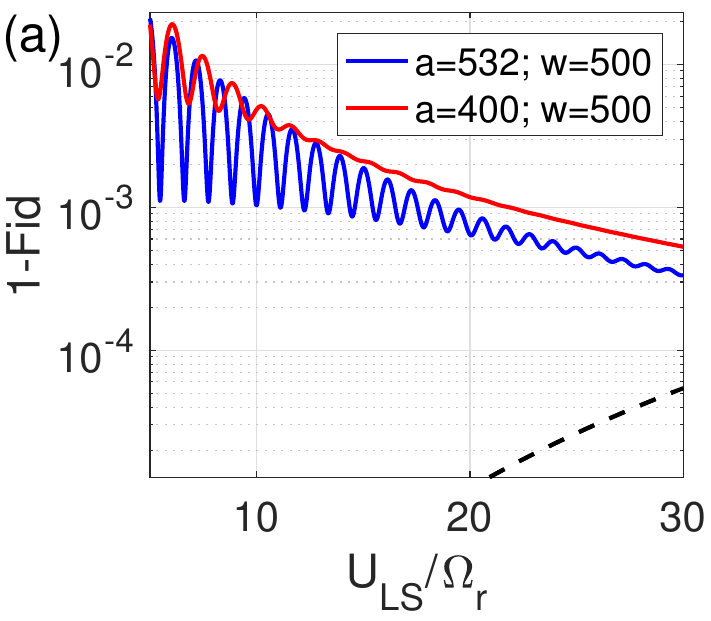}} 
       \scalebox{.35}{\includegraphics{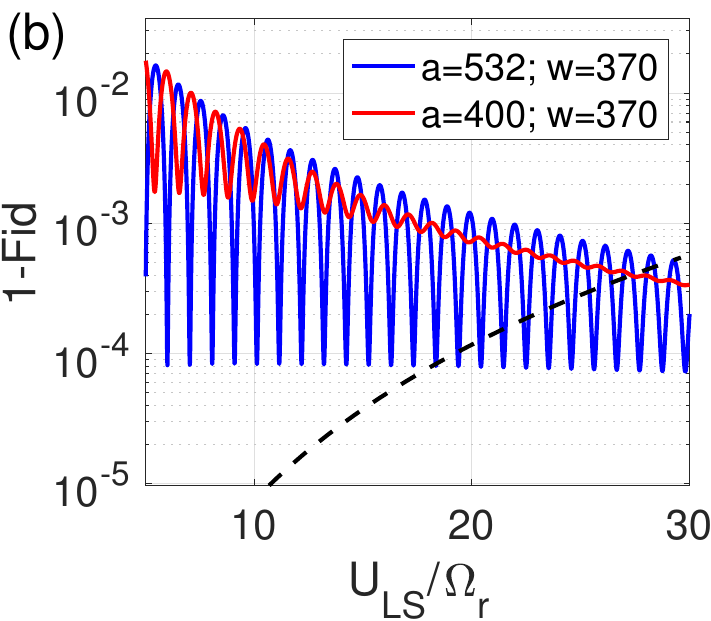}} 
\caption{Effects of Laser cross-talk and misalignment on gate fidelity. The population that does not return to the qubit basis after gate operation would be considered as loss. The errors are averaged over the laser's misalignment area which is a circle with radius $ r_0$  and also over the Wannier state of atoms. The reported error is averaged over all qubit configurations in a C$_4$-NOT gate. Red and blue lines are corresponding to systems described in Fig. 1 and 2 with lattice constants $a=400$ and  532 nm. The solid and dashed lines are corresponding to errors of plaquette $\bar{E}_p$ and central atoms $\bar{E}_c$. To apply the single-site addressing the 788nm laser is focused to $1/e^2$ intensity waist of (a) $w=$500nm and (b) $w=$370nm with alignment accuracy of $r_0=25$nm, generating a differential light-shift $U_{\text{LS}}$ on $|1 \rangle \langle r|$ transition. The Oscillation is due to the change of effective Rabi frequency $\tilde{\Omega}$, which leads to different values of Rydberg leakage at the plaquette sites after the gate operation. 
}\label{Fig_CrossTalk}
\end{figure}

  To quantify the single site addressing efficiency, we consider a microscope with NA=0.68 that focuses the 788nm light to $1/e^2$ intensity waists of $w=$370nm and 500nm \cite{Wei11}. The  alignment accuracy of 25nm  has been achieved for single-site addressing \cite{Wei11}, which is subject to improvement by e.g. sub-wavelength localization of atoms \cite{Mil13,Sub19}.  
 The generated light-shift by the focused laser has the form of $U_{\text{LS}}(x,y)=U_{\text{LS}} \text{e}^{-2((x-x_0)^2+(y-y_0)^2)/w^2}$ with $r_0=\sqrt{x_0^2+y_0^2}$ being the laser misalignment. At the central site, this misalignment would cause a detuning $\Delta(r_0)=U_{\text{LS}}(1-e^{-2r_0^2/w_0^2})$ in $\ket{1}\bra{r}$ transition that changes the effective Rabi frequency $\tilde{\Omega}=\sqrt{\Omega^2+\Delta^2}$. Hence for the central atom, starting at $\ket{1}$ qubit state, after the gate operation time $\tau=\frac{2\pi}{\Omega}$, the qubit state would not be fully retrieved, which results in an error of $E_c=\frac{\Omega^2}{\tilde{\Omega}^2}\sin^2(\tilde{\Omega}\tau/2)$.
 % from  and hinderes the retrive of qubit basis lead to population leakage after modifies the evolution and hence the final Rydberg population $P_r=\frac{\Omega^2}{\tilde{\Omega}^2}\sin^2(\tilde{\Omega}^2\tau/2)$ where $\tau=\frac{2\pi}{\Omega}$ is the duration of 2$\pi$ pulse.  
 %Small misalignment $\Delta(r_0)\ll \Omega$ results to the infidelity of $E_{m}(r)\approx\sin^2({\tilde{\Omega}^2\tau/2})$. 
 Considering the uncertainty in addressing a specific site within distance $r_0$, the  error of the centred atom must be averaged $\bar{E}_{c}=\frac{1}{\pi r_0^2}\int\limits_0^{r_0} E_{c}(r)2\pi r \,\text{d}r$. Taking into account the error distribution profile, the optimum operation time would be modified from $\tau=\frac{2\pi}{\Omega}$  to $\tau_{\text{opt}}=\frac{2\pi}{\sqrt{\Omega^2+\Delta(\frac{3}{4}r_0)}}$. The averaged error of the centered atom $\bar{E}_{c}$ is plotted by dashed lines in Fig.~\ref{Fig_CrossTalk} as a function of $U_{LS}/\Omega$. The infidelities caused by laser misalighnment could be suppressed by spatial beam shaping \cite{Gil16}.   
   
     \begin{figure*}
  \centering 
         \scalebox{.4}{\includegraphics{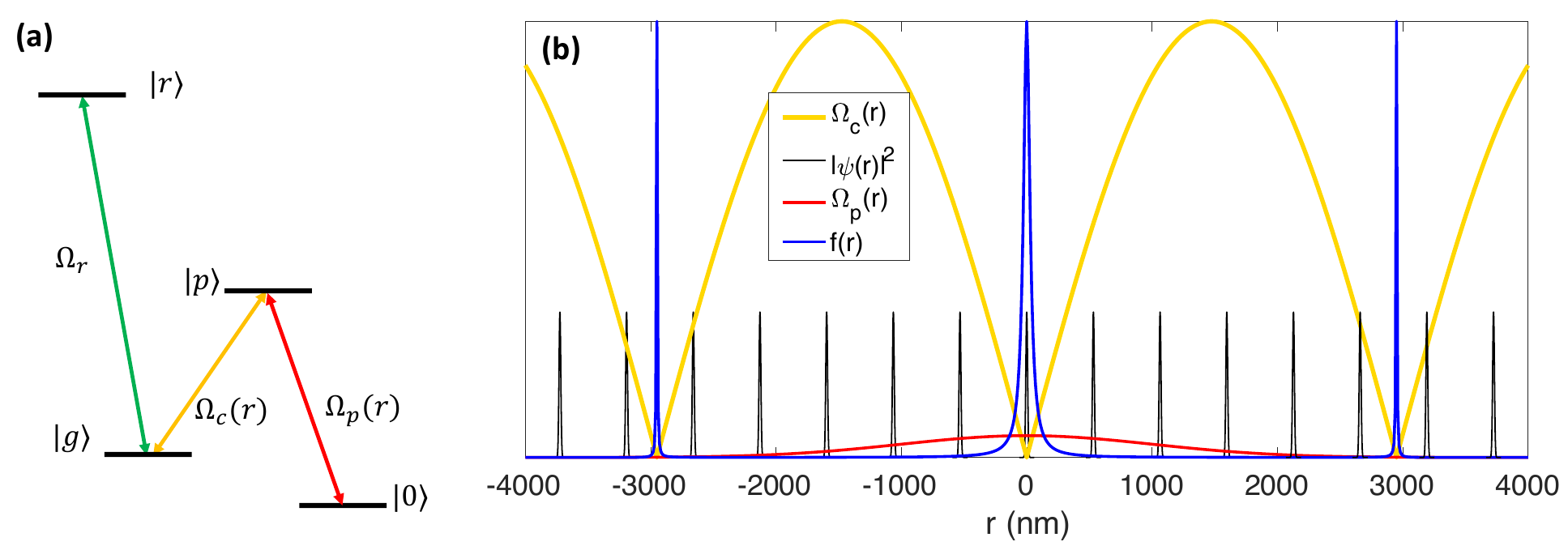}} 
\caption{Dark state single site addressing technique. 
(a) the level scheme containes a $\Lambda$ configuration transferring the population from $|0\rangle$ to $|g\rangle$ at the nodes of $\Omega_c$ standing wave where  $\Omega_c\ll\Omega_p$. The transferred atom would then get excited to the Rydberg state. (b) The spatial profile of the transition Rabi frequencies $\Omega_{c,p}$ as well as the $|0\rangle\langle g|$ transition probabilities $f(x)$ and the wave-function density $|\psi(x)|^2$ of the atoms in $|0\rangle$ qubit states  are plotted. $f(r)$ maps the $|\psi(x)|^2$  to $|g\rangle$ state upon the spatial overlap which is designed to only occurs at the targeted site. 
In (b) sample applied parameters are $\Omega_c/\Omega_p=30$, $\tilde{k}=1\mu m^{-1}$ and the Gaussian width of $\Omega_p$ laser is $w=2\mu$m. 
}\label{Fig_DarkAddress}
\end{figure*}  

Concerning the laser cross talk, the detuning experienced by the neighbouring plaquette atoms must be large enough to avoid population leakage out of the qubit basis. The leakage probability of a plaquette atom in $\ket{1}$ qubit state over the operation time $\tau_{\text{opt}}$ is given by
 %Over the Rydberg excitation of central atom, the probability of exciting a neighboring plaquette atom  in $\ket{1}$ qubit state to $\ket{r}$ over the operation time $\tau_{\text{opt}}$ is given by  
   $E_{p}=\frac{\Omega^2}{\Omega^2+\Delta(|{\bf r}|)^2}\sin^2(\frac{\Omega^2+\Delta(|{\bf r}|)^2}{\Omega^2+\Delta(\frac{3}{4}r_0)^2}\pi)$ where $|{\bf r}|=|{\bf a}-{\bf r_l}-{\bf r_p}|$ is the distance from the centre of the laser beam to the neighbouring plaquette atom. Here ${\bf a}$ is the distance vector between centered and a  plaquette site, and one needs to average over the laser misalignment ${\bf r_l}$ and also over the position of the plaquette atom ${\bf r_p}$ considering its Wannier wave-function to find the average loss population  $\bar{E}_p$ of a plaquette atom in $\ket{1}$ state.
The variation of the averaged plaquette error is plotted by solid lines in Fig.~\ref{Fig_CrossTalk}  %Averaging the above-mentioned errors over all qubit configurations, the effects of imperfections in single-site addressing  over the C$_4$-NOT gate operation are quantified in Fig. \ref{Fig_CrossTalk} 
as a function of $U_{\text{LS}}/\Omega$. The Oscillation is due to the change of effective Rabi frequency $\tilde{\Omega}$, which leads to different values of Rydberg leakage at the plaquette sites after the gate operation. At large $U_{LS}/\Omega$ and also for weak laser  focusing (large $w$) the variation of detuning over the plaquette atoms' wave-functions would be large and hence averaging the error over ${\bf r}_1$ and ${\bf r}_p$ washes the oscillation pattern. The laser cross talk could be suppressed by using two species lattices \cite{Huf22,Khaz22Log,She21,Sin21}.
Considering both plaquette error $\bar{E}_p$ and the central atom error $\bar{E}_c$, Fig.~\ref{Fig_CrossTalk} shows that single-site operations with high fidelity is achievable in the designed setups discussed in the main text.% As an example for $U_{LS}/\Omega_r=15$ and 
%Please note that using the auxiliary laser to make the individual site light-shift improves the accuracy in single site addressing. It also 
%Fig.~\ref{Fig_CrossTalk}a shows the leackage after the gate operation averaged over qubit configurations. Fig. b shows the average population of Rydberg atoms over the gate operation. 

%  \begin{figure}[h]
 % \centering 
   %    \scalebox{.39}{\includegraphics{CrossTalkAverageed}} 
%\caption{Effects of Laser cross-talk and misalignment. The errors are averaged over the misalignment area that is a circle with radius $ r_0$  and over the wave function of the ground motional state of trapped atoms. The reported error is averaged over all qubit configurations in a C$_4$-NOT gate. Dashed and solid lines are corresponding to systems described in Fig. 1 and 2 with lattice constants $a=400$ and  532 nm. The 788nm Laser is focused to $1/e^2$ width of $w=$500nm with alignment accuracy of $r_0=0.05a$, generating a differential light-shift $U_{\text{LS}}$ on $|1 \rangle \langle r|$ transition. The Oscillation is due to the change of effective Rabi frequency $\tilde{\Omega}$, which leads to different values of Rydberg leakage at the plaquette sites after the gate operation.
%}\label{Fig_CrossTalk}
%\end{figure}  

The scattering of the auxiliary laser could also affect the fidelity. To give an example a 420nm auxiliary laser focused to 370nm, dresses the $\ket{1}$ qubit state by $\ket{6P_{1/2}}$ with $\Omega_{LS}/2\pi=200$MHz and the laser detuning of $\Delta_{LS}/\Omega_{LS}=-150$. This laser imposes a differential light-shift of $U_{LS}=\Omega_{LS}^2/4\Delta_{LS}=-2.1$MHz on the $\ket{1}- \ket{r}$ transition.
In a Toffoli gate with $\Omega_r/2\pi=30$kHz (Fig. 4), the single-site addressing infidelity of 0.003 is expected, see Fig.~10b. Over the $30\mu$s operation time, the photon scattering from the $\ket{6P_{1/2}}$ state would cause 0.0015 gate infidelity.
%Using a stronger laser and going closer to resonance allows faster operation desired for the fan-out gate. For example focusing the 5mW laser \cite{Lev18} to 370nm waist results in $\Omega_{LS}/2\pi=3.3$GHz. Going closer to resonance $\Delta_{LS}/\Omega_{LS}=-30$, the differential light-shift would be $U_{LS} =-170$MHz for the  $\ket{1}- \ket{r}$ transition. For $\Omega_r=17$MHz,  the single-site addressing infidelity of 0.003 would be obtained, while the scattering infidelity would be as low as 0.00004.
An alternative approach is to initially change the hyperfine state of the desired site \cite{Wei11} and then excite the new auxiliary hyperfine state to the Rydberg level. In this case, it would be important that both hyperfine states get trapped at the same position in the qubit-dependent lattice of Fig.~11a-c. A possible choice is changing $\ket{0}=\ket{F=1,m_f=1}$  to $\ket{F=2,m_f=-1}$ which has the same distribution of $m_j$ components and hence experiences the same trapping potential. 

 \subsection{Interferometric approach}

 Single site addressing could be realized with precisions below the diffraction limit using an interferometer technique applied before in sub-wavelength localization \cite{Sub19,Kap10,Mil13,Cho07,Aga06}. 
Over this process the qubit state $\ket{0}=\ket{5S,F=1,m_f=1}$ of the desired site would be changed to an auxiliary hyperfine state  $\ket{g}=\ket{5S,F=2,m_f=-1}$  via an intermediate level $\ket{6P}$. The three-level $\Lambda$ transition is operated by a standing-wave driving field ($\Omega_c$)  and a focused laser  ($\Omega_p$), see Fig.~\ref{Fig_DarkAddress}. The standing wave is formed in each dimension by counter-propagating fields $\Omega_{c1q}\exp(ikq)$ and $\Omega_{c2q}\exp(-ikq+\phi_q)$ where  $q\in\{x,y\}$.
 The transition occurs under the dark state STIRAP mechanism.  The  dark-state in the described $\Lambda$ system is a superposition of the $\ket{g}$ and $\ket{0}$ states with spatially varying amplitudes: 
  \begin{equation}
  \ket{D({\bf r})}=\frac{1}{\sqrt{ \Omega_c({\bf r})^2+\Omega_p({\bf r})^2}}[\Omega_c({\bf r})\ket{0}-\Omega_p({\bf r})\ket{g}].
  \label{Eq_dark}
  \end{equation}
To apply the transition, first the $\Omega_{c1q}$ field would be applied. 
The probe field would then be applied focused on the targeted site with a Gaussian profile $\Omega_p({\bf r})=\Omega_p \text{e}^{-({\bf r}-{\bf r_0})^2/w^2}$ and $\Omega_p\ll\Omega_c$. In the next step $\Omega_{c2q}$ would be applied adiabatically \cite{Sub19} to form the  standing-wave with a node being adjusted on the position of the targeted site via the $\phi_{2q}$ angle.  
At the nodes of $\Omega_c$ standing-wave $\Omega_p(r) \gg \Omega_c(r)$  the dark-state composition is predominantly $\ket{g}$ while away from the nodes $\Omega_p(r)\ll \Omega_c(r)$ the dark state would remain at $\ket{0}$ qubit state.

Considering the wave-function density $|\psi({\bf r})|^2$ of atoms initialized in $\ket{0}$ states,  the local population of $\ket{g}$ state after applying $\Omega_{p,c}$ fields would be $f({\bf r})|\psi({\bf r})|^2$ where $f({\bf r})$ is obtained from Eq.~\ref{Eq_dark} as 
  \begin{equation}
  f(r)=\frac{\Omega_p^2 \text{e}^{-\frac{2({\bf r}-{\bf r_0})^2}{w^2}}}{\Omega_p^2 \text{e}^{-\frac{2({\bf r}-{\bf r_0})^2}{w^2}}+\Omega_c^2 \sin^2\tilde{k}(x-x_0)\sin^2\tilde{k}(y-y_0)},
  \end{equation}
where $\tilde{k}=k\sin \theta/2$ with $\theta$ being the angle between the $\Omega_{c1}$ and $\Omega_{c2}$ lasers. Figure~\ref{Fig_DarkAddress}b plots the  narrow peaks of  $f(r)$ at the nodes of $\Omega_c$. Going away from the focusing point of $\Omega_p$ at ${\bf r_0}$, the profile width of $f(r)$ gets narrower and  disappears. The full width at half maximum of an $f$ peak located at $r'$ would be given by $\text{FWHM}_{f(r')}=2\Omega_p\exp(-({\bf r'}-{\bf r_0})^2/w^2)/\tilde{k}\Omega_c$ \cite{Mil13,Sub19}.
%The nodes of $\Omega_c$ are at the incommensurate distance of the optical lattice sites and hence only at the aligned targeted site the presence of atomic population leads to the transition to $\ket{g}$. 
While the nearest peaks of $f$ shown in Fig.~\ref{Fig_DarkAddress}b do not overlap with the atomic lattice sites the next nearest neighbors are at the position where the amplitude of $\Omega_p$ would approach zero.
%the next nearest peakes of $f(r)$ would not have overlap with the lattice sites and at further distances, the amplitude of $\Omega_p$ would approach zero.
 In the next step, the $\Omega_p$ and $\Omega_c$ lasers would be turned off simultaneously keeping the ratio of $\Omega_c(t)/\Omega_p(t)$ constant to preserve the dark state components. At this stage, only the desired site would be in the $\ket{g}$ state and hence  would get  excited to the Rydberg level by the subsequent $\Omega_r$ laser. Considering the qubit-dependent lattice of Fig.~1b,  the  auxiliary state $\ket{g}$ would experience the same trapping potential as $\ket{0}$ state since the distribution of $m_j$ components in the two hyperfine states are the same.

The initial calibration of the $\Omega_c$ standing wave with the optical lattice could be done by fluorescence imaging with the approach of \cite{Sub19}. The nodes of the $\Omega_c$ standing wave could be then moved by high resolution adjusting of the $\Omega_{c2q}$  phase \cite{Tel20}.
%Then ?c2 beam is turned on with a different phase ?c2 implemented by changing the phase of the rf drive to the acousto-optic modulator (AOM), with the amplitude being ramped up to 250 ? with the optimal waveform so as to preserve the adiabaticity during the ramp. By scanning ?c2 from 0° to 360°, we change the position of the node of ?cðxÞ?500?cosðkxÞ, thereby mapping out the probability amplitude of atoms in each spatial slice of the wave function.
%Phase resolutions of $\text{d}\phi=2^{\circ}$ \cite{Tel20} results to the node alignment precision of $r_e=7.6$nm in an $\Omega_c$ standing-wave with the node separation of $1.38\mu$m.
For the chosen parameters of Fig.~\ref{Fig_DarkAddress}b applied in a lattice of $a=532$nm with the atom confinement of FWHM=20nm, and focusing $\Omega_p$ laser to the Gaussian width of 2$\mu$m, the single site addressing infidelity averaged over the qubit configurations would be 0.01. This calculation encounters the population leakage of the neighboring lattices as well as the imperfect transition of the targeted site. Using the two species lattice could improve the addressing fidelity. %The same parameters in a two species lattice suppresses the infidelity to 0.007.

 \subsection{Appendix~F: Laser excitation of the motional states in the Optical Lattice}

The laser excitation of atoms to the Rydberg state could lead to phases that depend on the atomic position. This could excite the motional states in the optical lattice. Let us consider the targeted atom in electronic and motional state $|1_e,0_m\rangle$. The spatial variation of the two-photon excitation with the counter-propagating 1013nm and 420nm lasers is given by $e^{ik\hat{z}}$ with $k=k_{1013}-k_{420}$.
% Applying the part of the Hamiltonian that is due to the laser to this state gives $(\Omega_{r}\ e^{ik\hat{z}_j}|r\rangle\langle 1|)|1\rangle |0\rangle_m=\Omega_{r}|e\rangle_je^{ik\hat{z}}|0\rangle_m$.
We can rewrite the vibrations of the position operator as $\hat{z}=\sigma/2(\hat{a}_j^{\dagger}+\hat{a}_j)$, where $\sigma=\sqrt{\frac{\hbar}{m\omega_{tr}}}$ is the spread of the ground motional state wave-function, $\omega_{tr}$ is the trap frequency and $(\hat{a},\hat{a}^{\dagger})$ are the phononic annihilation-creation operators of the targeted site. In the Lamb-Dicke regime $(\eta=k\sigma/2\ll1)$  one can expand the exponential to get
\begin{equation}
 e^{ik\hat{z }}=(l+i\eta(\hat{a}+\hat{a}^{\dagger} )+O(\eta^2)).
\end{equation}
The Hamiltonian describing the laser excitation can now be written in the new basis $|1_e,0_m\rangle, \, |r_e,0_m\rangle,\, |r_e,1_m\rangle$ as:
\begin{equation}
\left(\begin{array}{ccc}
0 & \Omega_{r} & \eta\Omega_{r}\\
\Omega_{r} & 0 & 0\\
\eta\Omega_{r} & 0 & \omega_{tr}
\end{array}\right)\left(\begin{array}{c}
|1_e,0_m\rangle\\
|r_e,0_m\rangle\\
|r_e,1_m\rangle
\end{array}\right)
\end{equation}
Considering the setups described before Eq.~\ref{Eq_64Sup} and Eq.~\ref{Eq_75Sup}  with FWHM$_z$=35nm,  the probability of exciting a motional state
$ |r_e,1_m\rangle$ over the Toffoli and fan-out operations with $\Omega_r/2\pi=30$ kHz and 30MHz would be 0.3\% and 1.5\% respectively.

\section{App. G: Alternative encoding of the qubit states}

\begin{figure}[h]
  \centering 
         \scalebox{.38}{\includegraphics{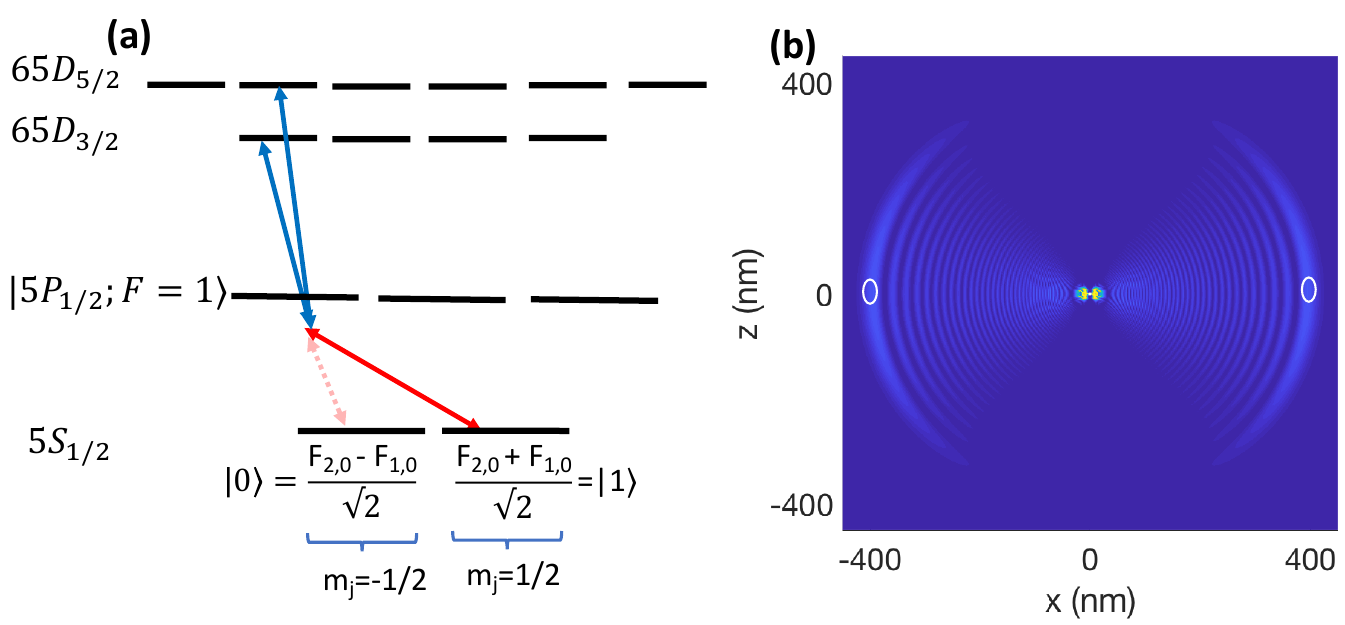}} 
\caption{Qubit encoding in the Hadamard combination of long-lived hyperfine states $|F=2,m_f=0\rangle$ and $|F=1,m_f=0\rangle$, see Eq.~\ref{Eq_encoding}. The two-qubit states have distinct $m_j$ components allowing qubit-dependent trapping, see Fig.~\ref{Fig_SpiLat}a-c. (a) Applying a two-photon Rydberg excitation via  $\sigma^{-}$ circular polarized lasers that are red detuned from the $|6P_{1/2};F=1,m_f=-1\rangle$ intermediate state, only the $|1\rangle$ qubit state would get excited to the Rydberg level, see the main text. The space-dependent Rydberg-Fermi interaction of the excited Rydberg level $(|64D_{3/2},-3/2\rangle+|64D_{5/2},-3/2\rangle)/\sqrt{2}$ is plotted in (b) where $z$ is perpendicular to the lattice plane.}
\label{Fig_StableEncoding}
\end{figure}

While the chosen qubit states have been widely used in the quantum information experiments in spin-dependent and -independent lattices \cite{Wei11,Man03,Lev18}, the dual encoding of the qubit could also be realized in other hyperfine states with longer coherence times.
%scheme could also be realized sensitivity of  these hyperfine states to the magnetic field affects the coherence time. 
%There are two approaches for addressing this problem. In the first approach one can store the qubits in the long lived hyperfine states $\ket{F=2,m_f=0}$ and $\ket{F=1,m_f=0}$ and only transfer the operating qubits to the new lattice sensitive qubit states $\ket{F=2,m_f=2}$ and $\ket{F=1,m_f=1}$ via three $\pi$-pulse dipole transitions generated by micro-wave fields with polarizations $\pi\sigma^{+}\sigma^{+}$  and return them afterwards by three pulses with polarizations $\sigma^{-}\sigma^{-}\pi$. Single site addressing has been realized in \cite{Wei11}.
%An alternative approach is encoding the information in the superposition states:
One example is the Hadamard combinations of the long-lived hyperfine states $\ket{F=1,m_f=0}$ and $\ket{F=2,m_f=0}$:
\begin{eqnarray}
\label{Eq_encoding}
 \ket{0}&=&(\ket{F=2,m_f=0}-\ket{F=1,m_f=0})/\sqrt{2} \quad \quad \\ \nonumber
 &=&\ket{I=3/2,m_I=1/2}\ket{J=1/2,m_j=-1/2} \quad \quad \\ \nonumber
 \\ \nonumber
 \ket{1}&=&(\ket{F=2,m_f=0}+\ket{F=1,m_f=0})/\sqrt{2} \quad \quad \\ \nonumber
  &=&\ket{I=3/2,m_I=-1/2}\ket{J=1/2,m_j=1/2} \quad \quad \\ \nonumber
 \end{eqnarray}
 In this arrangement the qubit states $\ket{0}$ and $\ket{1}$ would exclusively contain $m_j=-1/2$ and $m_j=1/2$ respectively. Hence they would get trapped by different polarizations of the qubit-dependent lattice as discussed in Fig.~\ref{Fig_SpiLat}a-c.
 In a Rydberg two-photon excitation with $\sigma^{-}$ circularly polarized lasers that are red detuned from the $\ket{6P_{1/2}}$ intermediate state, only the $\ket{1}$ state would get excited to the Rydberg level as discussed below.  Dipole transitions between the hyperfine states are given by
 \begin{eqnarray}\label{dipole}
\begin{array}{c} 
\bra{n'l'j';F'M'}\vec{r}\ket{nlj;FM}=
(-1)^{1+l'+s+J+J'+I-M'}\\
\sqrt{\text{max}(l.l')}\sqrt{(2J+1)(2J'+1)(2F+1)(2F'+1)} \\
\left\{ \begin{array}{ccc}
l'& J' & s\\
J & l & 1
\end{array}\right\}
\left\{ \begin{array}{ccc}
J' & F' & I'\\
F & J & 1
\end{array}\right\}
\left( \begin{array}{ccc}
F & 1 & F'\\
M & q & -M'
\end{array}\right) \langle n'l' |r| nl \rangle
\end{array}
\end{eqnarray}
where $q=0, \pm1$ for the linear  $\sigma^{0}$ and  $\sigma^{\pm}$ circular polarizations of the exciting light. Under the $\sigma^{-}$ circularly polarized laser  $\bra{6P_{1/2};1,-1}\vec{r}\ket{5S_{1/2};1,0}=\bra{6P_{1/2};1,-1}\vec{r}\ket{5S_{1/2};2,0}$, hence the dipole transition from the $\ket{0} $ ($\ket{1}$) qubit states of Eq.~\ref{Eq_encoding} to the  $\ket{6P_{1/2};F=1,M=-1}$ intermediate state would be forbidden (allowed) due to destructive (constructive) interference, see Fig.~\ref{Fig_StableEncoding}a.

In the upper transition of Fig.~\ref{Fig_StableEncoding}a, a two-color transition excite a superposition of the Rydberg levels $(\ket{64D_{3/2},-3/2}+\ket{64D_{5/2},-3/2})/\sqrt{2}$.  The two-color laser could be obtained in a setup of beamsplitters and acusto-optical modulators. The spatial profile of the Rydberg-Fermi interaction relative to the position of plaquette atoms is plotted in Fig.~\ref{Fig_StableEncoding}b.
The generated Rydberg superposition state mainly contains the $Y_{2,-2}$ spherical harmonic term, which  concentrates the electron wave-function close to the lattice plane and  enhances the interaction strength. The plaquette atoms experience an effective Rydberg-Fermi interaction that is averaged over their spatial profile. 
The scattering energy of Rydberg electron over the qubit-dependent Wannier state of  the l$^{th}$ plaquette atom in the geometry of Fig.~1 with FWHM$_{x,y}$=20nm and FWHM$_{z}$=35nm would be quantified by Eq.~\ref{Eq_V01} as
\begin{eqnarray}
\label{Eq_64D}
 &&\bar{V}_{\text{RF}\ket{1_l}}= 1.1\text{MHz},\quad \text{MD}_{V_{\text{RF}\ket{1_l}}}=0.14\text{MHz}\\ \nonumber
 &&\bar{V}_{\text{RF}\ket{0_l}}=0.27\text{MHz}, \quad \text{MD}_{V_{\text{RF}\ket{0_l}}}=0.16\text{MHz} \quad \quad
 \end{eqnarray}
where the in-plane qubit-dependent lattice-shift of $D=34.5$nm is considered. The same Rydberg state in the geometry of Fig.~2   reduces the unwanted level shift of the $\ket{0}$ qubit state to 
\begin{eqnarray}
\label{Eq_64D}
 &&\bar{V}_{\text{RF}\ket{0_l}}=0.16\text{MHz}, \quad \text{MD}_{V_{\text{RF}\ket{0_l}}}=0.09\text{MHz} \quad \quad
 \end{eqnarray}
with the qubit-dependent lattice shift of $D_z=150$nm being perpendicular to the lattice plane.

% \section*{Appendix~E: Laser alignment}
 
 %In [Nature 471, 319 (2011)] single site addressing in a lattice of a_lat=532nm is performed by focusing the UV light with NA=0.68. The position of the focused light could be adjusted by less than 10nm accuracy. The position of the sites are determined by 0.05alat accuracy which would be 20nm and 25nm in the scheme of Fig.1 and Fig.2 of our paper. This error is subject to improvement by enhanced resolution of CCD and tightly trapping of atoms. The same experimental setup has recently performed single site Rydberg excitation [PRL 128, 113602 (2022)]. 

%The 25 nm is mainly limited by the precision of site location. This is subject to enhancement by dark state technique in measuring the position of atoms [](dark state trapping to further localized atoms and) using ion microscope [] to detect positions of few reference sites for calibration of the laser. After that their laser has precisions below 10nm in addressing single sites. 


\begin{thebibliography}{100}
\bibitem{BackPhase} The background phase on each target atoms i.e.  $\tau \bar{V}_{\text{RF}\ket{0_l}}$ and the $\pi$ phase generated over the Rydberg excitation on the $\ket{1_c}$ state should be compensated by the laser light-shifts addressing individual sites. Alternatively, the central atom could be excited by dark state approaches to avoid the phase acquisition.
\bibitem{Alternative}The same order of lattice constant could also realized by the conventional $6S-6P$ transition via dark state approach \cite{Sun11} or by using dual species that are node and antinode seekers. Also the subwavelength confinement could be arranged using the dark state approach inspired by \cite{Yan18,Aga06}. The recently proposed twist-optical-lattice could confine atoms down to  sub-nanometer scale \cite{Kha23}.
\end{thebibliography}
\end{document}